\documentclass[global,twocolumn]{svjour}
\usepackage{graphicx}
\journalname{Journal of Materials Science}

\begin{document}
\title{Self-assembled nanowires on semiconductor surfaces}
\author{J.H.G.Owen\inst{1}\thanks{email:james.owen@nims.go.jp}\thanks{http://homepage.mac.com/jhgowen/research/research.html} \and
  K.Miki\inst{1}\thanks{email:miki.kazushi@nims.go.jp} \and 
  D.R.Bowler\inst{1,2,3}\thanks{email:david.bowler@ucl.ac.uk}%
}                     
\institute{International Centre for Young Scientists and Nanomaterials Laboratory, NIMS, 1-1
  Namiki, Tsukuba, Ibaraki 305-0044, JAPAN  \and
  Department of Physics \& Astronomy, University College London, Gower
  Street, London WC1E 6BT, UK \and London Centre for Nanotechnology,
  University College London, Gower Street, London WC1E 6BT, UK}
\date{Received: date / Revised version: date}
%
\maketitle
\begin{abstract}
  A number of different families of nanowires which self-assemble on
  semiconductor surfaces have been identified in recent years.  They
  are particularly interesting from the standpoint of nanoelectronics,
  which seeks non-lithographic ways of creating interconnects at the
  nanometre scale (though possibly for carrying signal rather than
  current), as well as from the standpoint of traditional materials
  science and surface science.  We survey these families and consider
  their physical and electronic structure, as well as their formation
  and reactivity.  Particular attention is paid to rare earth
  nanowires and the Bi nanoline, both of
  which self-assemble on Si(001).
\end{abstract}

\section{Introduction}
\label{sec:introduction}
  
As the scale of architectures in integrated circuit design continues to be reduced, 
the dimensions of present-day interconnects are on the scale of tens
of nanometers, and the 
interconnects for next-generation nanoelectronic devices may well be just
a few nanometers in lateral dimension. Furthermore, the incorporation of
nanometer-scale components, such as active molecules, into integrated circuits
will require interconnects of a similar scale. On these scales, self-assembled 
nanowire systems are becoming increasingly interesting\cite{Bowler2004,Barth2005}, 
as conventional lithographic techniques reach their limits around 10-15\,nm\cite{Chen1993} 
and SPM-based nanolithography\cite{Hashizume1996} methods lack scalability.
Moreover, the introduction of  such ``bottom-up'' technology, based on naturally 
nanometer-scale components to complement or even replace the current 
``top-down'' technology is likely to require completely different architectures.  
One example is the ``crossbar'' architecture\cite{Heath1998}, in which active 
molecules are used as devices at the junctions between two perpendicular 
nanowires.  This architecture is designed to take advantage of the typical 
product of self-assembly schemes -- an array of parallel wires -- rather than 
relying on controlled positioning of individual wires.
Over the last ten to fifteen years, a number of systems have been identified which 
yield self-assembled nanowires on semiconductor surfaces, and stem from a mixture
of serendipity and deliberate design. 
There are only a few systems which provide long, straight nanolines on flat terraces.  
In/Si(111)\cite{Nogami1987} (amongst other metals) and Pt/Ge(001)\cite{Gurlu2003} are
examples of systems which undergo surface reconstructions leading to 1D 
atomic chains with metallic properties.  On the Si(001) surface, the Bi nanolines\cite{Naitoh1997,Miki1999a} appear to be the only example of an atomically-perfect, 
straight, self-assembled 1D nanostructure, but are not metallic.
Aside from fortuitous discoveries, symmetry-breaking by use of a vicinal surface to produce
and array of steps can be used to grow long, straight nanowires of many metals, 
including Au, Bi et al.\cite{Baski2001}, and 1-D nanoscale structures can be designed
by taking advantage of anisotropy in the heteroepitaxial strain between the Si(001) 
surface and an appropriate deposited material. The rare-earth silicides are the archetype 
of this method\cite{Preinesberger1998}. Within the family of
rare-earths (as well as scandium and yttrium), a close lattice match can be obtained in one direction, with a
variety of positive and negative lattice mismatchs in the orthogonal
direction.  By choosing the right material, metastable 1D nanoscale
structures 3-10~nm wide and hundreds of nanometres long can be formed.
These have some phenomenological similarities with the Bi nanolines,
as will be described.  This review will survey the growing field of
nanowires on semiconductor surfaces, with the focus on the Bi/Si(001)
nanoline system, whose importance is explained in the next section.

The study of semiconductor surfaces has been revolutionised over the
last twenty years by the advent of two techniques which give
real-space atomic scale information: Scanning Tunneling Microscopy
(STM) and electronic structure modelling (in particular, Density
Functional Theory, or DFT).  The two techniques are complementary, in
that STM gives an approximate answer to an exact question (i.e. the
structure of the real Si(001) surface), while DFT gives an exact
answer to an approximate problem (i.e. the lowest-energy structure of
a computation cell containing a few Si dimers, and a few layers of
bulk-like Si).  While these techniques between them enabled the clear
identification of the Si(001) reconstruction, even today
there is some controversy about the exact nature of the buckling of
dimers on Si(001) at low temperatures\cite{Sagisaka2003}.  In the same
way, the understanding of the structure and properties of the
nanolines surveyed in this paper has been greatly enhanced by a close
collaboration between experiment and theory. The benefits of this
synergistic approach will be drawn out throughout the discussion.

\subsection{Introducing the Bi nanolines}
\label{sec:BiIntro}

\begin{figure}
\includegraphics[width=\columnwidth]{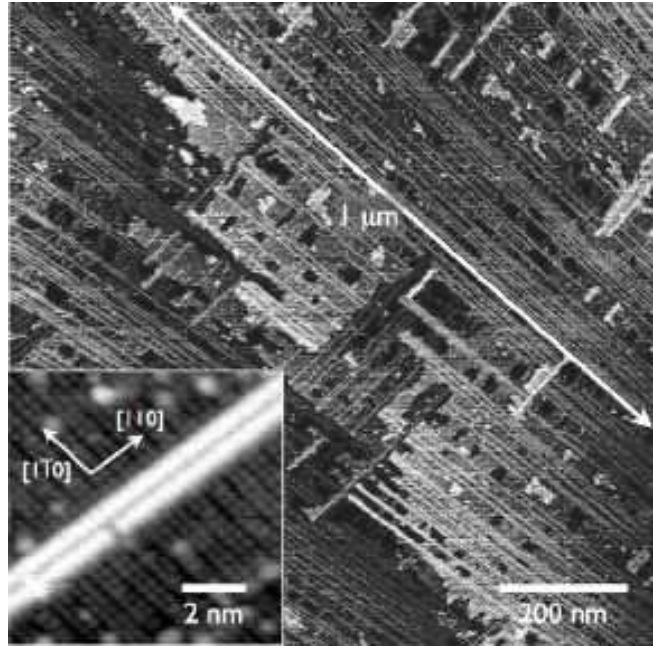}
\caption{A typical high-temperature STM image of the Bi/Si(001)
  surface. Here 1$\mu$m$\times$1$\mu$m. The Bi nanolines run 
  along $\langle110\rangle$ directions, perpendicular to the Si dimer rows, 
  and can grow to over 1$\mu$m in length.  In the inset, the atomic structure, 
  comprising a pair of Bi dimers, about 6.5\AA\ apart, is shown. Hydrogen 
  termination of the Bi nanolines has been used to enhance the 
  apparent height of the nanoline. }
\label{fig:Binanoline}
\end{figure}

The Bi:Si(001) nanoline system, which will be the focus of this
review, was discovered about ten years ago.  As can be seen from the
examples in Fig.~\ref{fig:Binanoline}, they grow perfectly straight
along the $\langle110\rangle$ directions on the Si(001) surface for
hundreds of nanometres, apparently limited only by the terrace size of
the underlying substrate. When they encounter a step edge, they will
either grow out over lower terraces in a long, narrow promontory, or
they will burrow into higher terraces, until a deep inlet is formed.
They have a constant width, 1.5~nm or four substrate dimers, and are
very stable. So long as the temperature is not so high that the Bi can
evaporate, they are stable against prolonged annealing, maintaining
the same width.  This is quite unlike the rare earth silicide systems
which in most cases (Yttrium being the probable exception) will coarsen into 3D
epitaxial islands with long anneals.  There are no other elements
which produce perfect nanolines on the Si(001) surface, although there
is some evidence to suggest that Sb nanowires with the same appearance
as the Bi nanoline form in the presence of surface
hydrogen\cite{Kubo1999}, and DFT modelling that indicates that a Sb
nanoline would be stable\cite{Wang2003}.  There has also been a
suggestion that Er may form nanolines of the same
structure\cite{Wang2005} (as opposed to silicide wires), but this
remains unconfirmed.

Structurally, the Bi nanoline is unique amongst nanoline systems---it
is neither a periodic reconstruction of the surface, nor the result of
anisotropically strained heteroepitaxial growth of one bulk structure
on another.  In fact, it is somewhere in between. The nanoline
structure is built around a pair of Bi ad-dimers, on top of a complex
subsurface reconstruction which is responsible for many of the
nanoline's remarkable properties, such as the extreme straightness;
this will be explained in detail below. In this article, we intend to 
describe the formation, structure and properties of this system, 
setting it in the context of other nanoline systems, 
particularly the rare-earth silicide family.

\subsection{Overview}
\label{sec:overview}

The paper is laid out as follows: in Section~\ref{sec:methods} we
consider the methods used to examine the systems, both experimental
and theoretical; in Section~\ref{sec:other-nanowires} self-assembled
nanowire systems \emph{other} than the Bi:Si(001) nanowire system are
described, with particular reference to the rare-earth silicide
family; in Section~\ref{sec:bi-nanowires} the Bi:Si(001) nanowire
system is examined in detail, and in Section~\ref{sec:conclusions} the
state of the field is considered and conclusions are drawn.  In
considering the nanowires in Sections~\ref{sec:other-nanowires}
and~\ref{sec:bi-nanowires} we consider first the formation and
structure of the systems (including the effects of annealing),
followed by the electronic properties (including conductance data
where available) and finish with the reactivity of the systems.

\section{Methods}
\label{sec:methods}

\subsection{Hot STM}
\label{sec:hot-stm}

Since its invention in 1982 by Binnig and Rohrer at IBM, STM has allowed 
researchers to image dynamic surface processes, in real-space, with 
near-atomic resolution. 
STM as a technique bridges a variety of disciplines. In surface physics, 
STM is unique as a real-space probe of electronic density of states of 
individual species; the STM tip can be a passive observer of
surface chemical reactions, and also an active catalyst of these
reactions; and STM allows for the direct observation of many phase 
transformation processes, albeit in 2D rather than 3D. 
However, this is not meant to be a review of capabilities of STM; for that, 
the reader is pointed to classic books on the subject\cite{JChen1993,Wiesendanger1994}.
Here a few comments are made, which are pertinent to the particular challenges
of elevated-temperature (`hot') STM for the study of epitaxial growth.

Surface reactions can be followed at room temperature by using quench experiments, in which
a high-temperature surface is cooled rapidly to freeze in a snapshot of
an evolving surface, for example in the growth of III-V semiconductor surfaces\cite{Avery1997}. 
However, the quenching process may introduce spurious surface features, and so it
is preferable to image the surface \textit{in-situ}, i.e. at temperature.
The ability to vary the sample temperature makes it
possible to image a surface in the process of changing, rather than after it has
changed.  To give just one example, the phase transition between the (1$\times$1) 
reconstruction and the (7$\times$7) reconstruction on the Si(111) surface, at around
1100K, has been observed directly in both directions,
by careful control of the temperature across the transition\cite{Miki1992}.  
For useful observations of dynamic processes, the speed of the reaction must be 
matched to the rather slow imaging rate of an STM.  One example is the 
motion of step edges\cite{Swartzentruber1996}. At around 300$^{\circ}$C, step motion can be imaged
as the addition or subtraction of 4 atoms between STM images. However, above about
500$^{\circ}$C, the step edge will have moved significantly in
the time taken for the STM to scan one line. Thus from line to line,
the step will be in different places, as it oscillates about a mean
position. In STM, therefore, fast-moving steps have a streaky
appearance above 500$^\circ$C, as may be seen in Fig.~\ref{fig:nucleationSTM}. Furthermore, 
mobile species may be invisible at elevated temperatures, either due to the 
decrease in contrast, or due to their mobility. 

Surface chemical processes, such as the diffusion of Si atoms 
on Si(001)\cite{Swartzentruber1996} or the decomposition of a 
molecule\cite{Rezaei1999}, can be studied
at the single-species level, and complete chemical reaction pathways from
initial adsorption  can be followed through to complete 
decomposition\cite{Wang1994,Owen1997b}.  Kinetic information such as 
activation barriers can be extracted, and any metastable intermediate 
structures can be imaged, giving a unique insight into the reaction process.  
The ability to control the substrate temperature is
particularly important in studies of epitaxial growth, where the
kinetic pathways available at different temperatures will determine
the surface morphology. In the case of Si/Si(001) and Ge/Si(001)
epitaxy, it has been possible to image the surface simultaneously with
the deposition of material, and therefore observe directly the
nucleation of islands, and the growth of these nuclei into 1D, 2D and
3D islands, along with associated processes, such as the ripening of
epitaxial islands, surface stress relief mechanisms, and the
transition from island growth to step flow growth
modes\cite{Voigtlander1997,Owen1997c,Goldfarb1997b}, 
all of which processes occur at different sample temperatures.

However, these added abilities bring with them a penalty in
sensitivity.  There is generally a loss of resolution due to thermal
noise above room temperature, and a loss of stability, with thermal
drift making it hard to image the same area for extended periods.  In
many of the high-temperature STM images shown here, the nanolines have
a curved appearance. They have no curvature themselves, the appearance
comes from the drift during scanning.  Furthermore, imaging of
epitaxial growth necessarily involves imaging a surface where there
are many atoms loosely bound to the surface, or in the gas phase being
deposited. In this situation, it is very easy for atoms to stick to
the STM tip, which can have adverse effects on the imaging, and it is
usually preferable to obtain stable imaging conditions rather than
achieve the highest possible resolution, which is generally a more
unstable situation.  For the same reason, Scanning Tunneling
Spectroscopy (STS), or Current Imaging Tunneling
Spectroscopy(CITS)\footnote{A process in which a rapid voltage ramp is
  applied to the STM tip while in a fixed position above the surface,
  generating an I/V curve from which the LDOS can be measured},
becomes unfeasible at elevated temperatures.  It is generally better
to take a series of images of the same area at different bias
voltages, giving voltage contrast of an object, although as the
feedback loop is active in this situation, the tip-sample distance
will vary, and so the information gained is not equivalent to
STS/CITS.  Thus these limitations mean that while high-temperature STM
brings the unique ability to study evolving surface morphology in
real-space, it is not always the best way to study growth surfaces. A
combination of high-temperature studies, and room-temperature studies
of quenched surfaces, is likely to provide a fuller description of a system.

\subsection{Electronic Structure Modelling}
\label{sec:modelling}

There have been a number of books on modelling techniques
published recently\cite{Martin2004,Finnis2004}, so this section will
feature only a brief description of the techniques used for modelling
of semiconductor surfaces, concentrating on electronic structure
calculations (both semi-empirical and \textit{ab initio}).  These
techniques retain the quantum mechanics of the electrons while
treating the ions classically.

The first problem to consider is how to model a semi-infinite piece of
material (that is, a surface).  The approximation that is used is to
consider a small piece of appropriate material with boundary
conditions.  Generally, one of two solutions is used: either a cluster
of atoms, with the dangling bonds on the edges terminated in some
suitable way (e.g.  hydrogen), or a slab of atoms with periodic
boundary conditions in two or three directions.  A cluster allows a
smaller number of atoms to be used than a slab, but has the drawback
that it is hard to allow for long-range strain effects (such as those
caused by reconstructions or steps on surfaces).  A slab approach
requires care: the vacuum gap between repeating images must be large
enough to prevent interactions, and the slab itself must be
sufficiently thick to allow strain relaxation.  The slab approach is
more commonly used, particularly with the methods described below.

Once the system to be modelled has been defined, the calculation
technique must be chosen.  Putting aside quantum chemical methods for
the sake of brevity (they scale rather poorly with system size---for
more details, the reader is directed to one of many excellent books on
the subject\cite{Szabo1989}), we will consider the semi-empirical tight
binding (TB) technique, and the \textit{ab initio} density functional
theory (DFT) technique.

TB\cite{Goringe1997,Ordejon1998} postulates a basis set of
atom-centred orbitals\footnote{For the class of methods known as
  \textit{ab initio} tight binding\cite{Sankey1989,Horsfield2000}, the
  basis is explicit.}  for the wavefunctions; the Schr\"odinger
equation can then be rewritten as a matrix equation, with elements of
the Hamiltonian matrix formed from integrals between orbitals on
different atoms.  These Hamiltonian matrix elements are \emph{fitted}
to either \textit{ab initio} calculated data or experimental results.
For simplicity, the basis is often assumed to be nearest neighbour
only (with a cutoff defined on the range of interactions) and
orthogonal.  Once the Hamiltonian has been defined, the band energy of
the system can be found by diagonalisation of the Hamiltonian or other
methods.  The other energetic terms (e.g.  Hartree correction,
exchange and ion-ion interactions) are represented by a repulsive
potential (often a pair potential).

Despite its apparent simplicity, TB is extremely effective, and
generally qualitatively accurate, if not approaching quantitative.  It
has been shown\cite{Sutton1988} that this TB formalism can be derived
from DFT via a set of well-defined approximations, which explains to
some extent its success.  It is well-suited for a rapid exploration of
configuration space (e.g. possible structures for some new surface
feature\cite{Bowler2002}) provided that a suitable parameterisation exists for the
bonds between different species.  The fitting of parameterisations is
non-trivial, and the transferrability of a given parameterisation
(i.e. its accuracy in environments far from those in which the
fitting was performed) is never guaranteed.  Generally the matrix
elements themselves are fitted to a band structure (or energy levels
of a molecule), while their scaling with distance is fitted to elastic
constants or normal modes of the system.  TB retains quantum mechanics
(since the energy is obtained by solving the Sch\"odinger equation)
while approximating the most complex problems.

DFT\cite{Hohenberg-kohn-1964,Kohn-sham-1965} starts with an exact
reformulation of the quantum mechanics of a system of interacting
electrons in an external potential: the result is a set of equations
for non-interacting electrons moving in an \emph{effective} potential
with all the complex electron-electron interactions in a single term,
known as the exchange-correlation functional (it is a function of the
charge density, which is itself a function of position, hence
``functional'').  Unfortunately, the form of this functional is not
known, and must be approximated, for instance with the local density
approximation (LDA) or one of the generalised gradient approximations
(GGA).

DFT has been extremely successful in many areas of physics, chemistry,
materials and, increasingly, biochemistry, particularly when combined
with the pseudopotential approximation.  The hard nuclear potential
and the core electrons of each atom are replaced with a single
``pseudopotential'', and only the valence electrons are considered,
leading to a softer potential.  This approximation is most effective
for atoms where there is considerable screening of the valence
electrons by the core electrons (e.g. in Si the 3s electrons are
screened by the 1s and 2s shells, while the 3p electrons are screened
by the 2p shell); in first row elements and first row transition
metals, this is a much smaller effect, leading to the development of
``ultrasoft'' pseudopotentials.  These issues are discussed in more
detail in many excellent books and
reviews\cite{Lindan2002,Martin2004}.  The essential point to note is
that DFT is widely used for electronic structure calculations, and
provides results which are accurate to within approximately 0.1\,eV.

An important development in electronic structure techniques over the
last ten years is that of linear scaling
techniques\cite{Goedecker-1999}.  Standard techniques for solving for
the ground state, whether TB or DFT, scale with the cube of the number
of atoms in the system (either because matrix diagonalisation scales
with the cube of the matrix size, or because the eigenfunctions spread
over the whole simulation cell, leading to cubic scaling when they are
orthonormalised).  This scaling places rather strong restrictions on
the sizes of system which can be modelled, even on massively parallel
machines: for \textit{ab initio} methods, going beyond 1,000 atoms
rapidly becomes prohibitive, though somewhat larger systems can be
addressed.  However, since electronic structure is fundamentally local
(consider bonding as an example), the amount of \emph{information} in
the system should be proportional to the number of \emph{atoms} in the
system.  Tight binding methods which take advantage of this (e.g.
refs.\cite{Li1993,Goedecker1994}) have been widely used for some time,
allowing calculations on many thousands of atoms.  The implementation
of linear scaling DFT algorithms has proved significantly harder,
though recent efforts in nearly linear\cite{fattebert00} and linear
scaling methods\cite{Ozaki2001,soler02,Bowler2002b,Skylaris05} suggest
that these techniques are starting to produce useful and general results.

\subsection{Modelling STM}
\label{sec:modelling-stm}

One of the challenges of using STM to examine semiconductor surfaces
is that the current arises from both geometric and electronic
structure, though this is less true at high biases (where height or
geometric structure will dominate).  Some technique is required to
understand these changes, and to test proposed structures against
experiment.  The field of STM simulation is a complex one (in part
because the structure and composition of the tip is unknown); the
interested reader is referred to excellent reviews for further
information\cite{Briggs1999,Hofer2003}.  We will briefly summarise the
simplest and most common approximation used, and discuss how the
experimental-theoretical interaction can best proceed.

The most commonly used approach is the Tersoff-Hamann
approximation\cite{Tersoff1985}.  This asserts (via a careful series
of approximations) that the tunneling current is proportional to the
local density of states (LDOS) due to the sample at the position of the
tip.  In effect, the tip is assumed to have a flat density of states.
This approach is directly equivalent to considering the projected
charge density (i.e. the partial charge density due to each band for
bands within $V_{bias}$ of the Fermi level).  It is qualitatively
accurate, though will not reproduce the observed corrugation values
correctly\cite{Hofer2003}. 

The authors are of the opinion that for true success in investigating
nanowires (and other systems) on semiconductor surfaces, a close
collaboration between experiment and modelling is required.  Any
successful collaboration depends on a number of factors, but the key
factors, we believe, are frequent, easy correspondence, an
understanding of the limitations and abilities of the appropriate
techniques, and trust between the different sides in the
collaboration.  The last point extends to sharing of unpublished data,
and extensive discussion of possible courses of action.  The three
authors have been working together in different combinations for ten
years (leading to around 20 joint publications), and we find that our
collaboration is still immensely fruitful.

\section{Other Nanowires}
\label{sec:other-nanowires}

While the Bi/Si(001) nanolines have remarkable structural qualities,
they are not the only self-assembled nanowire system on semiconductor
surfaces.  In this section we present an overview of these other
nanowire systems (more details are presented
elsewhere\cite{Bowler2004}), and we will consider in more detail the
most important of these systems, the rare-earth silicide family,
focussing on their structure and formation, electronic properties and
reactivity.

\begin{figure}
  \centering
  \includegraphics[width=\columnwidth]{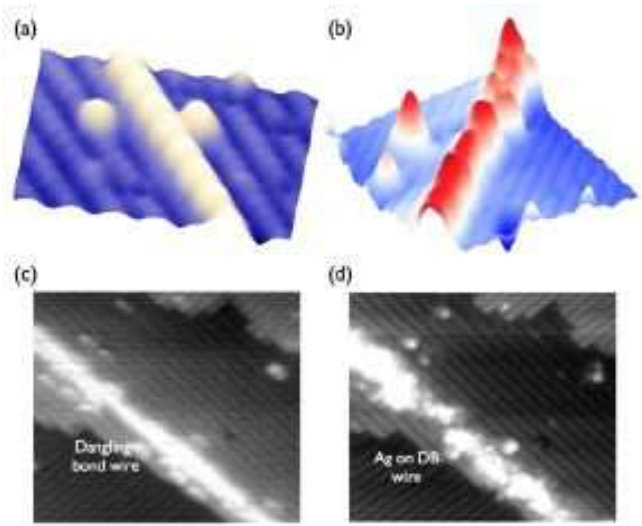}
  \caption{(a): A dangling-bond wire on the H:Si(001) surface. (b) Ga deposited along a row of dangling bonds on H:Si(001). (c): A line of Si dimers after removal of H termination by an STM tip. (d): Preferential adsorption of Ag onto the clean Si dimers, to form an atomic-scale wire. (a,b) courtesy of T. Hashizume, Hitachi ARL, Hatoyama, Saitama, Japan.  (c,d) courtesy of T.Sakurai, NIMS, Tsukuba, Ibaraki, Japan.}
  \label{fig:Hashizume}
\end{figure}

A variety of metals have been found to self-assemble into nanowires or
nanolines on the Si(001) surface, though none with the perfection of
the Bi nanolines. The simplest way in which to form 1D structures on
the Si(001) surface is by epitaxial growth. An island nucleus is
essentially an ad-dimer, which sits between the dimer rows. The two
pairs of Si dimers which support the ad-dimer are distorted by its
presence, and so are attractive adsorption sites for further ad-atoms
or ad-dimers\cite{Swartzentruber1997}. This process has been described
as a surface polymerisation reaction\cite{Brocks1993}.  The long sides
of the string, which are equivalent to A-type steps on this surface,
have a very low sticking coefficient which keeps the island from
broadening at lower growth temperatures. Deposition of Group III and
Group IV elements on Si(001) will therefore result in long 1D chains
of
dimers\cite{Evans1999,Dong2001,Takeuchi2001,Srivastava2004,Albao2005}.
However, there is little control over the length or structure of these
wires, and they are unstable against annealing, eventually
reorganising into compact islands.  The Pt/Ge(001)
system\cite{Gurlu2003} forms nanowires through a complex surface
reconstruction, resulting in a system with a high degree of
perfection, which approaches that of the Bi nanolines. When annealed to high
temperatures, arrays of atomically perfect nanowires of Pt/Ge form,
each of which is 0.4~nm wide with a spacing of 1.6~nm between the
wires.

Sideways growth of two-dimensional islands can also be blocked by
introduction of missing dimer trenches. These form to relieve surface
stress, either because of contamination by a small amount of a
transition metal such as Ni, or through heteroepitaxial growth of
another material such as Ge. The trenches tend to line up in a
semiregular array, giving an approximate (2$\times$n) reconstruction
in LEED, with n$\sim$8-12 (though the regularity of the trenches is
rather poor).  This method has been used to deposit a variety of
metals, such as Fe\cite{Kida1999}, Ga\cite{Wang2002} and
In\cite{Li2001a} (using the Ni-based technique) and the molecule
styrene\cite{Zhang2004}. The result is long-range, reasonably well
ordered wires of the metals, though their properties have not been
characterised.

The 2D symmetry of the surface can also be broken deliberately by the use of vicinal 
substrates to produce regular arrays of steps\cite{Lin1998} which can then be 
decorated with materials such as
gold\cite{Baski2001,Losio2001,Lee2002,Ahn2003,Crain2003} and
Bi\cite{Cho2004}. This approach is used primarily on Si(111), although with
an extreme orientation towards (112) directions the surface might be
labelled Si(557) or Si(5 5 12) depending on the angle.  The resulting step
reconstructions comprise chains which are metallic and one-dimensional; 
early observations suggested possible observations of Luttinger liquid
behaviour\cite{Segovia1999}, though this has been disputed\cite{Losio2001,Ahn2003}.  
Their spacing can be controlled to some extent by varying the angle of miscut of the surface.

As an atomic-scale extension of the idea of lithography, the Si(001)
surface has been passivated with atomic hydrogen (forming the
monohydride phase) and individual hydrogen atoms removed with an STM
tip to form atomic-scale patterns.  The wire formed by removal of
hydrogen atoms along (or across) a dimer row is known as a ``dangling
bond'' wire (or DB
wire)\cite{Shen1995,Watanabe1996,Watanabe1997,Hitosugi1997,Hitosugi1999,Bird2003a}
and is predicted to show conduction effects similar to conjugated
polymers with polaronic and solitonic
effects\cite{Bowler2001,Todorovic2002,Bird2003}.  An example of a
dangling-bond wire is shown in Fig.~\ref{fig:Hashizume}(a). These
wires have been reacted with a variety of adsorbates including
iron\cite{Adams1996}, gallium\cite{Hashizume1996,Hashizume1997},
aluminium\cite{Shen1997}, silver\cite{Sakurai2000} and organic
molecules such as norbornadiene\cite{Hersham2000,Abeln1998,Abeln1997}.
However, although for a single row of dangling bonds, a perfect atomic
wire can be formed, as with Ga in Fig.~\ref{fig:Hashizume}(b), for
greater widths, as in Fig.~\ref{fig:Hashizume}(c,d), the dangling bond wire is more ragged,
and the resulting wires (here Ag) are typically composed of a large
density of small, roughly spherical crystals, which exhibit many
imperfections and boundaries.  A more elegant implementation stems
from careful application of molecular chemistry to produce a directed reaction: 
when the monohydride surface was exposed to styrene and \emph{one} 
hydrogen atom removed, self-directed growth of
lines of styrene resulted from a chain reaction between each adsorbed
molecule and an adjacent hydrogen\cite{Lopinski2000}.  Such
self-organised methods are important for the growth of molecular
nanowires on surfaces.  However, for all these systems there is a lack
of control over the size and length of the nanowires.

\subsection{Structure and Formation of Rare-Earth Nanowires}
\label{sec:formation-structure}

\begin{figure}
  \centering
  \includegraphics[width=\columnwidth]{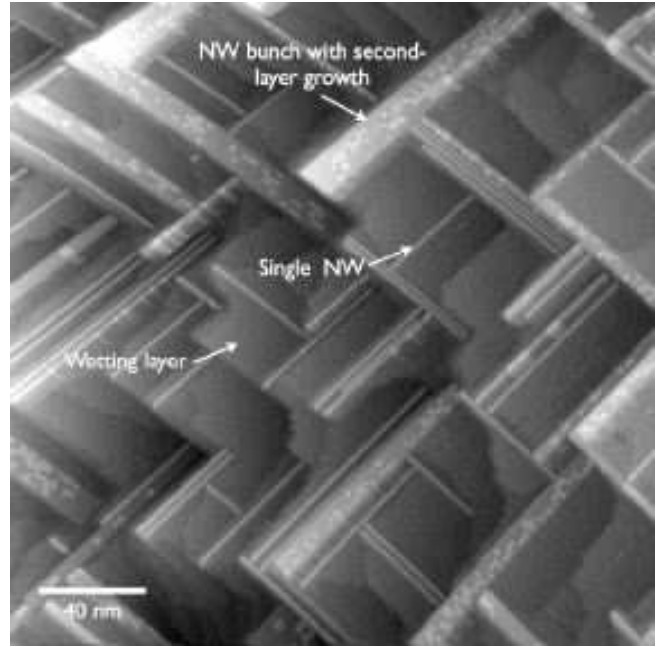}
  \caption{A large-scale image of Dy NWs, showing the 3 different types of surface feature
  associated with the RENWs. On the terraces, a reconstruction is typically seen, here (2$\times$7).
  Single NWs grow in both $\langle$110$\rangle$ directions, blocking each other's growth. Finally, 
  the NWs can form bunches. Both the single NWs and the bunches can grow a second layer of material,
  unlike the Bi nanolines. \textit{Image courtesy Prof. J.Nogami, U. Toronto, Canada.}}
  \label{fig:silicide1}
\end{figure}

A more promising materials system for nanowire growth on Si(001) 
is that of the rare-earth metal silicides.  There has been extensive research 
on these systems because of their good conductivity and low Schottky barrier with
silicon\cite{Netzer1995}, though this work concentrated on the Si(111)
surface.  Moreover, in the hexagonal phase some rare-earth
silicides show a good lattice match with the Si(001) substrate in one direction, with
a large mismatch in the other.  Heteroepitaxial growth of these materials would therefore 
be expected to be constrained in the high mismatch direction, and facile
in the other, resulting in long 1D islands. This is indeed what happens. 
The behaviour of the rare-earth nanowire systems on Si(001) has been 
investigated extensively, and nanowires have been observed under certain
growth conditions. As the other major family of self-assembled nanowires on the
Si(001) surface, we give a fuller description of this family for comparison
to the Bi nanolines. 

\begin{table}
  \centering
  \begin{tabular}{|l|ll|ll|}
  \hline
    Rare earth & a(\AA) & (\%) & c (\AA) & (\%)\\
  \hline
    ScSi$_{1.7}$ & 3.66 & (-4.69) & 3.87 & (+0.78)\\
    YSi$_2$ (hexagonal) & 3.842 & (+0.05) & 4.144 & (+7.92) \\
    Sm$_3$Si$_5$ (hexagonal) & 3.90 & (+1.64) & 4.21 & (+9.64) \\
    GdSi$_2$ (hexagonal) & 3.877 & (+0.96) & 4.172 & (+8.65) \\
    DySi$_2$ (hexagonal) & 3.831 & (-0.23) & 4.121 & (+7.32) \\
    HoSi$_2$ (hexagonal) & 3.816 & (-0.63) & 4.107 & (+6.95) \\
    ErSi$_{2-x}$ (hexagonal) & 3.79 & (-1.30) & 4.09 & (+6.51) \\
    YbSi$_2$ & 3.784 & (-1.46) & 4.098 & (+6.71) \\
  \hline
  \end{tabular}
  \caption{Lattice mismatch between hexagonal phase of various
    rare-earth silicides and Si(001).  Data is presented in terms of
    lattice constant and percentage mismatch (the Si(001) surface has
    a lattice constant of 3.84\AA).  Where the indication
    ``hexagonal'' is given, it indicates that other phases are
    possible.  \textit{Courtesy of Dr C. Ohbuchi, NIMS, Japan.}.}
  \label{tab:rare-earth-mismatch}
\end{table}

The most common form of rare-earth nanowire (RENW) formed on Si(001)
is thought to result from the anisotropic strain between the AlB$_2$
crystal structure of the nanowire and the Si(001) substrate, leading
to fast growth along the less-strained direction, and extremely
limited growth along the more-strained direction.  
The mismatch for various silicides in the AlB$_2$ hexagonal structure is
given in Table~\ref{tab:rare-earth-mismatch}. RENWs of this type
have been reported for
Er\cite{Chen2000,Chen2002b,Chen2002a,Chen2002,Ragan2003,Fitting2003,Lee2005,Harako2005},
Dy\cite{Preinesberger1998,Nogami2001,Preisenberger2002,Chen2002,Ragan2003,Liu2003b,Liu2003a,He2003,Liu2003,He2004b,He2004a,Lee2005},
Gd\cite{Chen2002,Liu2003b,Liu2003a,Lee2005,Harrison2005} and 
Ho\cite{Nogami2001,Ohbuchi2002}. For Sm\cite{Ragan2003,Lee2005} NWs only appear 
on vicinal substrates; other reports\cite{Nogami2001,Ohbuchi2005} find that on 
flat surfaces only rectangular, 3D islands are formed after a (2$\times$3) layer). Bundles of NWs have
been reported for Yb\cite{Kuzmin2004} ; again, early reports\cite{Nogami2001,Katkov2003}
suggested that NWs were not formed, and 3D islands resulted from
annealing.  Though they are not rare-earth metals, similar NWs have
been reported for both Sc\cite{Chen2002} and
Y\cite{Liu2003b,Katkov2002}, resulting from the same combination of
crystal structure and anisotropic strain.  Although it forms a similar
wetting layer on the substrate to Yb, Eu does not form
NWs\cite{Perala2005}.

The resulting RENWs are found both individually and in bundles; the
individual wires have sizes which vary from element to element but are
in general 5-10~nm wide and less than 1~nm high.  The wires grow
extremely fast, reaching lengths of up to 1$\mu$m; more interestingly
they always run along $\langle 110\rangle$ directions and will grow out over 
step edges, drawing terraces with them, features that they share with
Bi nanolines. The large-scale image in Fig.~\ref{fig:silicide1} is very similar to that
of Bi nanolines at the same scale.
 For almost all materials, the surface of the wires shows a c(2$\times$2)
reconstruction\cite{Chen2002b,Chen2002a,Ohbuchi2002,Liu2003a,Liu2003,He2004a},
though the Yb NWs, which are off-axis relative to the other RENWs, may
have a (1$\times$1) surface\cite{Kuzmin2004a}.

The Si(111) surface matches the lattice of the RE AlB$_2$ structure
reasonably well in all directions (with a mismatch of less than 2\%
for most metals)\cite{Netzer1995} and so would not be
expected to give anisotropic growth, but GdSi$_2$ NWs
have been formed along step edges of a vicinal
surface\cite{McChesney2002}.  These form because of a mismatch
perpendicular to the step edge, and grow to over 1$\mu$m long on
appropriately prepared samples, with a width of 10~nm and height of 0.6~nm.

Other NW structures have been reported for transition metals on
Si(111), for instance Fe\cite{Tanaka2005} (though these are rather
short wires), Ni\cite{Lin2004} and Ti\cite{Stevens2003,He2003a}.  The
growth mechanism for these wires is not entirely clear: the resulting
NWs are 5~nm high and 5~nm wide (NiSi$_2$), 10~nm high and 40~nm wide
(TiSi$_2$), though for both of these systems the NWs appear to grow
\emph{into} the substrate somewhat.  There is a technique for growing
NWs on Si(001), Si(110) and Si(111) explicitly relying on growth into
the substrate, resulting in ``endotaxial'' NWs, which has been applied
to Dy\cite{He2003} in the hexagonal AlB$_2$ structure, as well as
Co\cite{Okino2005,He2004b}, which takes the CaF$_2$ structure.  The
driving force appears to be kinetic in these systems: the long
direction of the NW has an interface which grows faster than the short
direction, though this is partly dependent on the structure of the
interface below the surface.

\begin{figure}
  \centering
  \includegraphics[width=\columnwidth]{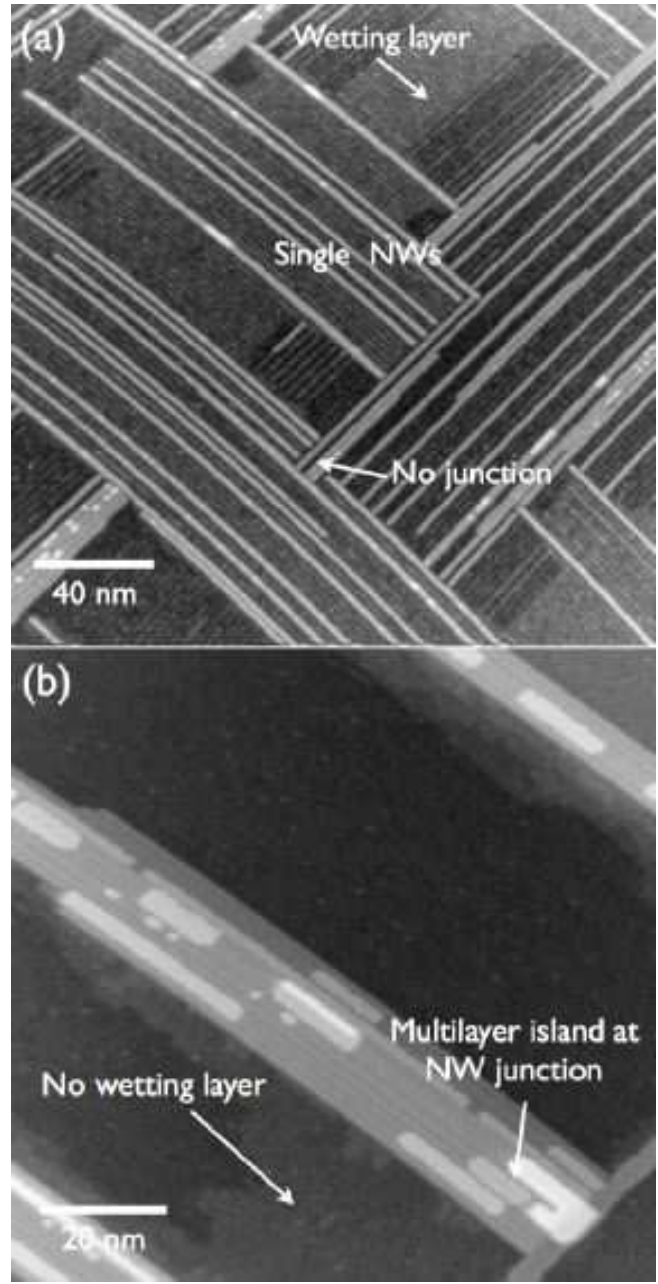}
  \caption{While a wetting layer is common for Dy silicide wires, it is
  not always present, and depends upon the total coverage, and annealing condition.
  In (a), there is a reconstructed wetting layer, and most NW are isolated and disconnected, 
  with a consistent width, although second-layer growth has occurred along some NWs. In (b), 
  the background is disordered Si(2x1), and most wires are bundles, which
  have joined together where they meet, forming multilayer islands. \textit{Images courtesy Prof. J.Nogami, U. Toronto, Canada.}}
  \label{fig:silicide2}
\end{figure}

The growth and formation of the RENWs is not understood at an atomic
level: they form extremely quickly (on experimental timescales) so
that, for instance, hot STM cannot be used to observe formation of the
lines.  Unlike the Bi nanolines, which have a constant width, and do
not change in width with annealing, the rare-earth nanoline family can
take a variety of widths depending on the growth conditions.  In the case of
DySi$_{2}$, single NWs of a consistent width form from the underlying (2$\times$7) 
reconstructed wetting layer, as in Fig.~\ref{fig:silicide2}. However, further 
annealing results in the formation of bunches of nanowires, which  
join together where they meet, and multilayer islands form.
As strained coherent heteroepitaxial islands, the RENWs are only metastable;
prolonged annealing will cause them to coarsen, and form large 3D
islands.  Overall, the deposition/anneal temperature must be in the
window between 550$^\circ$C and 650$^\circ$C, and annealing time
should not exceed 2-10 minutes.  Below 550$^\circ$C or 2 minutes, the
reaction will be incomplete and the NWs formed will be immature.
Above 650$^\circ$C or 10 minutes the systems tend towards lower energy,
thermodynamically stable states.  In particular, Dy-Si(001) will form
3D islands if annealed for long times (e.g. 30 minutes for
0.86ML)\cite{Liu2003} with wires formed for shorter annealing periods.
Annealing Er-Si(001) initially results in NWs, but as the annealing
time is increased dislocations form in the NWs\cite{Chen2002}.  If the
temperature is raised beyond 620$^\circ$C the same behaviour is seen,
leading ultimately to coarsening into islands\cite{Chen2002b}.  A
large-scale PEEM study of ErSi$_{2}$\cite{Fitting2003} found that large NWs
(which may have been bundles of nanowires---the technique does not
have the resolution to distinguish) were not affected by annealing on a
coarse scale.  However, Gd shows more complex behaviour. Annealing
studies of thin films of Gd found that both hexagonal and orthorhombic
phases could form\cite{Molnar1998}, and that with longer annealing
times the orthorhombic phase grew at the expense of the hexagonal
phase. Long anneals of the Gd/Si(001) system, up to 620$^\circ$C for
one hour, resulted in the formation of previously unknown silicide
structures, which are aligned perpendicular to hexagonal silicide NWs
on the same terrace\cite{Harrison2005b}.

There are various useful observations of their behaviour which can be
made:

\begin{itemize}
\item A wetting layer with a characteristic reconstruction forms
  before the nanowires, and persists on the substrate with the
  nanowires; specific reconstructions have been observed for:
  \begin{itemize}
  \item Eu-Si(001):(2$\times$3)\cite{Perala2005},
  \item Sm-Si(001):(2$\times$3)\cite{Ohbuchi2005}, 
  \item
    Yb-Si(001):(2$\times$3)\cite{Kuzmin2003,Katkov2003,Kuzmin2004,Perala2005} 
    and (2$\times$4)\cite{Katkov2003,Kuzmin2004},    
  \item Nd-Si(001):(2$\times$3) and (2$\times$4)\cite{Katkov2003},  
  \item Dy-Si(001):(2$\times$4)\cite{Liu2001}, and (2$\times$5) and
    (2$\times$7)\cite{Liu2003a},  
  \item Gd-Si(001):(2$\times$4)\cite{Liu2003a}, and (2$\times$5) and
    (2$\times$7)\cite{Liu2003a,Harrison2005}  
  \item Ho-Si(001):(2$\times$4) and (2$\times$7)\cite{Ohbuchi2002}
  \end{itemize}
\item The only system for which this does not happen is
  Er-Si(001)\cite{Chen2002b}. 
\item The system must be heated during deposition or annealed
  post-deposition to allow reaction of the rare-earth metal with the
  substrate\cite{Chen2000,Nogami2001}
\item Annealing the system for too long, or depositing at too high a
  temperature, can result in 3D islands rather than nanowires; this is
  discussed in more detail below.
\end{itemize}

As different metals have different properties for each of these
categories, we will discuss them briefly in turn.  Gd and
Dy\cite{Liu2003a} form a (2$\times$4) reconstruction first, followed
by a (2$\times$7) reconstruction\cite{Liu2003a,Harrison2005} just
prior to nanoline formation.  Rather similar behaviour is seem for
Ho\cite{Ohbuchi2002}.  The metal fraction in the two reconstructions
is 3/8 and 5/14 ML respectively\cite{Liu2001,Liu2003a}.  For these
three metals, the presence of the wetting layer is intimately
connected with formation and stability of the nanowires.

Despite the wealth of studies of ErSi$_2$ nanowires, there are no
reports of the structure and make-up of a wetting layer for this
system; there is some evidence\cite{Chen2002b} that once more than
$\sim$0.05ML of Er is deposited, long chains of surface dimers form,
followed by nanowires, with no intermediate reconstruction.  Indeed,
there are indications\cite{Ragan2005} that the surrounding substrate
is clean Si(001).

Another group of metals form closely related reconstructions:
Eu\cite{Perala2005,Kuzmin2005},
Yb\cite{Katkov2003,Kuzmin2003,Kuzmin2004,Perala2005},
Nd\cite{Katkov2003} and Sm\cite{Ohbuchi2005} all form (2$\times$3) and
in some cases (2$\times$4) reconstructions; by contrast to Er, Dy, Gd
and Ho, these metals are strained in both surface directions on
contact with Si(001) (which might be expected to reduce the
periodicity of any substrate wetting) and lack the main driving
force seen before for nanowire formation.  In general these metals do not form
nanowires\cite{Nogami2001,Katkov2003,Kuzmin2003} though there are
certain conditions where Yb can be made to form
nanowires\cite{Kuzmin2004a} of some kind, which may well be grown
off-axis (in a different growth mode to the other silicide nanowires
which grow with the c-axis aligned with the substrate).

It is also possible to induce nanowire formation in situations where
they would not normally form, either because of uniform lattice
matching (e.g on Si(111) surfaces, as mentioned above) or lattice
mismatch in \emph{both} surface directions using step edges.  This
technique has been successfully employed for Sm on
Si(001)\cite{Ragan2003,Lee2005} and Gd\cite{McChesney2002} and
Dy\cite{He2005} on Si(111).

\subsection{Electronic Properties of Rare-Earth Nanowires}
\label{sec:electr-prop}

One of the original technological interests in rare-earth overlayers
on Si(111) was the good crystal growth possible due to the interface
between the silicon and the overlayer, and the conductivity
properties.  The films have good conductivity and a low Schottky
barrier ($\sim$0.4 eV on n-Si and $\sim$0.8 eV on p-Si\cite{Netzer1995}).

Measurement of the conductivity of RENWs is a significant challenge.
Full two-point or four-point measurements requires either a unique STM
instrument (for instance one used to measure conductivity of CoSi$_2$
nanowires on Si(110)\cite{Okino2005}), or alternatively the formation of nanoscale 
contacts to a NW\cite{Fujimori2004}.  While significant progress has
been made in this area recently, any measurements
of conductivity will include the resistance of the interface between
the NW and the contacts.

Scanning tunneling spectroscopy (STS) has been performed on various
NWs: Dy and Ho\cite{Nogami2001,Ohbuchi2002} and
Gd\cite{Lee2003,Lee2005}.  While STS does not measure the conductivity
\emph{along} the NW, it does measure the local electronic structure,
and in all cases the NWs are found to be metallic, while the
surrounding reconstructed substrate is
not\cite{Nogami2001,Ohbuchi2002,Lee2005}.  This is good evidence that
the NWs are taking on the bulk silicide structure, which is conducting, and the
wetting layer is an intermediate state which relieves strain.

Conductivity measurements of transition metal silicides on Si(111) and Si(110)
have been made\cite{Lin2004,Okino2005}.  These studies exemplify the
two techniques for measuring NW conductivity: NiSi$_2$ NWs were
contacted by gold pads\cite{Lin2004}, while four-probe STM
measurements were made on CoSi$_2$ NWs\cite{Okino2005}.  These NWs are
relatively large compared to the RENWs discussed so far: 15nm (Ni) and
60nm (Co) wide.  The CoSi$_2$ NWs showed a high Schottky barrier and
conductivity equivalent to that of high quality thin films of
CoSi$_2$, while the NiSi$_2$ NWs showed some signatures of quantum
transport, though the conductivity of these wires was significantly
lower than thin films; this is likely due to the overgrowth of the
sample with SiO$_2$ for transfer to the lithography apparatus.

\subsection{Reactivity of Rare-Earth Nanowires}
\label{sec:reactivity-1}

There is very little data on the chemical reactivity of the rare-earth
nanowires.  Thin films on Si(111) are susceptible to reaction with
O\cite{Netzer1995}; the RENWs oxidise rapidly in air\cite{Ragan2005},
and transition metal NWs show signs of this (confirmed by the
conductivity of NiSi$_2$ NWs on Si(111)\cite{Lin2004}).  In the TM
system, the NWs were overgrown with native oxide, and showed a
significant decrease in conductivity relative to complete thin films
of similar thickness, which is attributed to scattering at the NW
surface/oxide interface.

Further information on the reactivity comes from the deposition of Pt
on a surface containing ErSi$_2$ NWs on Si(001)\cite{Ragan2005}.  The
Pt was deposited at room temperature and the sample was subsequently
annealed.  When STM images were taken they showed that the
c(2$\times$2) surface reconstruction on the NWs was no longer visible
though the (2$\times$1) substrate reconstruction was still present.
Furthermore, the Pt-covered NWs were resistant to reactive ion etching
and appeared stable when exposed to air for periods of up to 8 weeks;
the Pt overlayer does appear to strain the NWs, however, leading to
delamination of the NWs from the substrate if left untreated.  Clearly
there is much work to be done understanding the reactivity and
stability of these NWs.

\section{Bi Nanowires}
\label{sec:bi-nanowires}

The discovery that nanowires would form if a Bi-covered Si(001)
surface was left to anneal around the Bi desorption temperature was
made quite by chance.  In 1995, in the Materials Department of Oxford
University, two of the authors were investigating the surfactant-assisted
growth of Ge/Si(001) using Bi as a surfactant, and were studying the
properties of Bi/Si(001).  When a sample which had been left to anneal
at \textit{ca.}  500$^{\circ}$C overnight was imaged, very large, flat
terraces, and long, straight, bright lines (longer than the scanning
range of the STM), running across the surface were found.  These were
the Bi nanolines. The first published image of the ``nanobelts'' was
in 1997\cite{Naitoh1997}, and the first big study of their growth and
properties came in 1999\cite{Miki1999a}.  In this section, we will
discuss the atomic structure of the Bi nanoline, and its physical and
electronic properties. We show that many of the properties stem from
its unusual structure. The reaction of the Bi dimers with a variety of
reagents, and the effect of burial of the nanoline will also be
discussed.

\subsection{Physical Structure}
\label{sec:physical-structure}

\begin{figure}
\includegraphics[width=\columnwidth]{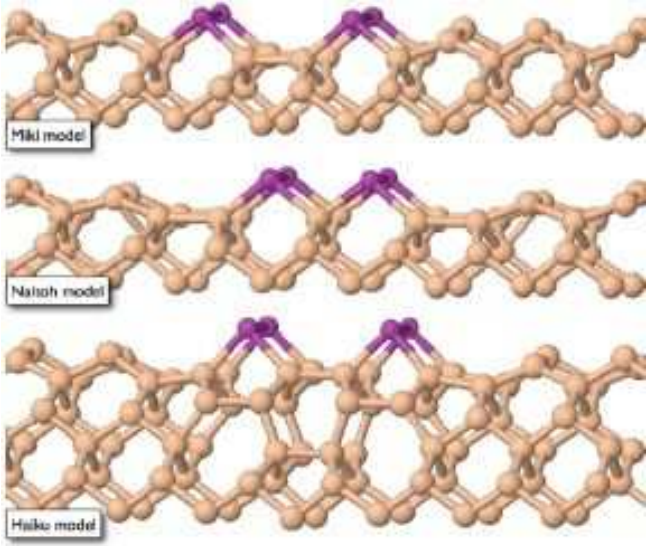}
\caption{Ball-and-stick models of the three proposed models for the Bi
  nanoline. (a) The Miki model.  \protect\cite{Miki1999a}.  (b) The
  Naitoh model\protect\cite{Naitoh2000}. (c) The Haiku
  model\protect\cite{Owen2002b}.}
  \label{fig:models}
\end{figure}

The physical structure is the key to understanding the properties of a
nanoline system, and great effort has been devoted to identifying the
structure of the Bi nanoline. In this case, the structure was
identified by a synergy of STM observations with tightbinding and DFT
simulations.  The three different proposed models for the Bi nanoline
are shown in Fig.~\ref{fig:models}.  The two early models (shown as
the top two models) share certain structural motifs. They contain a
pair of Bi dimers set into the surface layer of the Si(001) crystal,
whose compressive stress is relieved by missing dimer defects (DVs).
The first model proposed, shown in Fig.~\ref{fig:models}(a), was based
around a pair of Bi dimers in the surface layer, separated by a
missing dimer defect, so as to relieve the local stress of the Bi
dimers\cite{Miki1999a,Miki1999b}. This became known as the Miki model.
The second proposed model, shown in Fig.~\ref{fig:models}(b), was
based around a pair of Bi dimers with defects on either
side\cite{Naitoh2000}, which became known as the Naitoh model.
However, the third model is more complex: a reconstruction of several
layers of Si underneath the Bi dimers produces the nanoline core. This
was named the Haiku model\cite{Owen2002b}\footnote{It is named after
  the Japanese verse form, which comprises three lines of 5, 7, and 5
  syllables, referring to the 5- and 7-membered rings of silicon in
  the reconstruction.}.  Of the three models, this structure is the
only one to fit all the criteria which have been determined from STM
observations, as well as having the lowest energy of the three. In
this section, we will briefly review the historical process by which
this structure was determined, and discuss the properties of the Haiku
structure.

\subsubsection{Early Models}
\label{sec:early-models}
Early STM images of the Bi nanoline were taken at elevated
temperature\cite{Miki1999a,Miki1999b}, in which the Si dimers were not
resolved, and hence the registry of the nanoline with the substrate was
unknown. This data suggested that the nanoline contained two features,
which were probably Bi dimers, with a spacing of approximately 6.3\AA,
with a total width of about 1~nm, which was approximately equivalent to
the space of three Si dimers.  As the nanolines appeared bright at large
bias voltages (in both positive and negative bias images) to dark at low
voltages, it was also reasonable to assume that these Bi dimers were
situated in the surface layer, rather than in an adlayer or in a
subsurface layer. It was thought that Bi dimers embedded in the top
surface layer would have considerable compressive stress, and on this
basis, a model based around a 1DV, with Bi dimers to either side, was
proposed. This was the Miki model\cite{Miki1999a,Miki1999b,Bowler2000}.
Tightbinding calculations of the Miki model\cite{Bowler2000} found that
it was more stable than the (2$\times$1) or (2$\times$n) Bi reconstructions, and that
the formation energy of a defect in the line was high: around 1.1
eV\cite{Bowler2000} (this number considers putting the Bi which has been
removed as an ad-dimer on the surface; recent data\cite{Owen2005b} shows
that the defect energy falls to 0.11 eV if the Bi dimer is placed in
another Miki model; this point is discussed more fully in
Section~\ref{sec:miki-model-revisited} below).  DFT
calculations\cite{Miwa2002a,Owen2002b,Bowler2002,Miwa2002b} agreed with
those conclusions. The calculated LDOS showed that the Bi dimer states
were further away from the Fermi level than those of the Si dimers and
simulated STM images of the Miki
structure\cite{Miwa2002a,Miwa2002b,Srivastava2004} showed that the line
would appear dark at low bias voltages, in agreement with STM
results\cite{Owen2003,MacLeod2005}.  Testing by other methods appeared
to confirm this structure. Photoemission spectroscopy
experiments\cite{Owen2002b} found that the Bi 5d core-level spectra of
the Bi nanowire was essentially identical to the spectra of the
(2$\times$n) phase composed of Bi ad-dimers. This suggested that the
local chemical state and registry of Bi adsorbates for both phases was
the same, i.e.  that the Bi was in the form of dimers in the top layer
of the structure. X-ray photo-electron diffraction (XPD)
experiments\cite{Shimomura2000} found a good fit between the
experimental XPD data and simulated XPD intensity peaks from the Miki
model. In particular, they confirmed the presence of Bi dimers parallel
to the Si dimers and found the spacing between them to be $\sim$6.3\AA,
in agreement with the STM measurement, though it is important to note
that the quoted distances are the result of a \emph{fit} to the Miki model.

\begin{figure}
\includegraphics[width=\columnwidth]{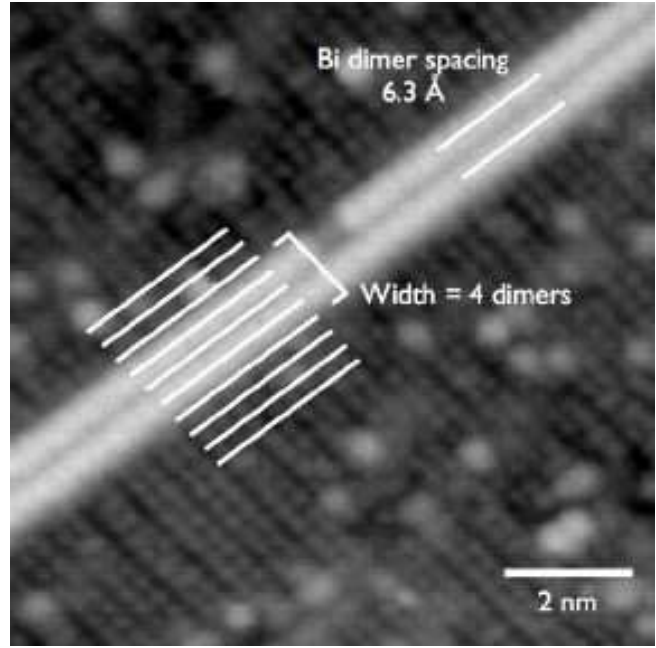}
\caption{10~nm$\times$10~nm STM image of the Si(001):H surface. The H
  termination makes it easier to resolve the individual Si dimers. A
  series of markers has been placed across the Bi nanoline. From this,
  it can be determined that the nanoline occupies the space of
  \emph{four} dimers.}
\label{fig:registry}
\end{figure}

However, it was difficult to account for the extreme straightness of
the nanoline with the Miki model, as the calculated kinking energy was
very small, around 0.1~eV. A kinetic argument was put forward, which
suggested that there was a stronger preference for incoming Bi dimers
to line up at the end of the nanoline, and hence the nanoline would
grow straight\cite{Miki1999a}. Furthermore, the diffusion constant of
a Bi surface dimer was high (this diffusion was required for a kink to
form in a Miki model nanoline), and hence a line would remain straight
once it had grown. Despite this explanation, the issue of the
straightness of the Bi nanolines remained as a question mark over the
Miki model.

Subsequent room-temperature STM experiments\cite{Naitoh2000} proved
decisively that the Miki model could not be the structure of the Bi
nanoline, as it had the wrong registry with the surface. Markers were
placed on the surface of an image with the Si dimers resolved, as shown
in Fig.~\ref{fig:registry}, and by counting across the nanoline, the
width was determined definitively to be four dimers, and not three as in
the Miki model; moreover the Bi dimers lay \emph{between} the Si dimers
of the substrate, not in the same position.  (This registry has been
confirmed by every other STM measurement
made\cite{Owen2002a,Owen2002b,MacLeod2004a,Wang2005,Miwa2005,Miwa2005b,Owen2005b}).
On the basis of the new STM data, the Naitoh model was
proposed\cite{Naitoh2000}.  Again this model was tested by atomistic
calculations; a detailed study of the dimensions and simulated STM
appearance\cite{Miwa2002b,Srivastava2004,Miwa2005b} showed that again
the nanoline would appear dark in STM at low bias voltages, but showed
also that the spacing of the Bi dimers was too narrow, approximately
5\AA, which could not be reconciled with the spacing of 6.3\AA\ measured
from STM. Moreover, the energy was considerably worse than the Miki
model\cite{Owen2002b,Miwa2002b}, and the kinking energy was still low.
There was therefore no satisfactory structural model at this time.

\begin{figure}
\includegraphics[width=\columnwidth]{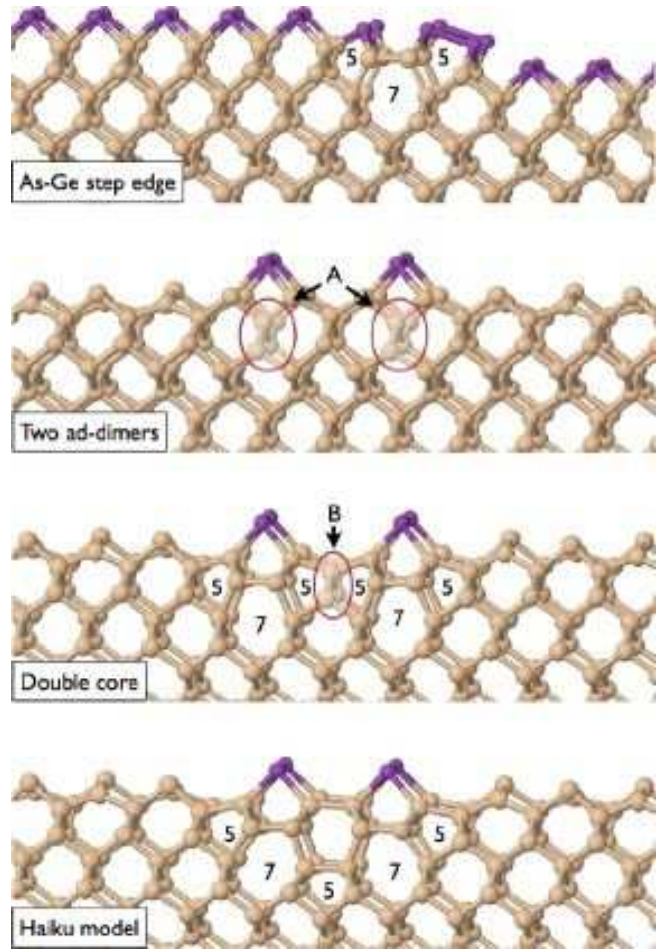}
  \caption{Ball-and-stick models of the As/Ge step edge structure, a
    pair of ad-dimers, the double-core structure, and the Haiku structure.
    Schematically, a double-core structure may be
    achieved by rotation of the atoms marked `A' underneath the ad-dimers from a vertical 
    plane into a horizontal plane. Removal of the 4 central atoms marked `B' and rebonding 
    then produces the Haiku structure. The 5-7-5 structural motif is also marked.}
\label{fig:haikucartoon}
\end{figure}

\subsubsection{The Haiku model}
\label{sec:haiku}
\begin{figure}
\includegraphics[width=\columnwidth]{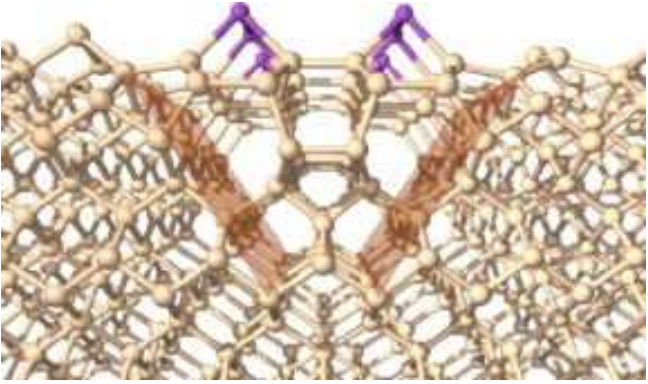}
  \caption{A cross-sectional perspective view of the Bi nanoline. The nanoline is 1.5~nm wide, 
  occupying the space of 4 surface dimers, and is built around a pair of Bi dimers, on top of a 
  complex Si substructure. The shaded \{111\} planes define the interface between the nanoline 
  core and the silicon substrate.}
\label{fig:haikucore}
\end{figure}

Inspiration for the structural model of the Bi nanoline arose from the
structural model proposed for the B-type double-height step of
As-terminated Ge\cite{Zhang2001}.  At this step edge, bond rotation
forms a pair of 5 and 7-membered rings of Ge, capped with As, shown in
Fig.~\ref{fig:haikucartoon}(a).  While the Haiku structure appears to
be a complex reconstruction, in fact the construction from a cell
containing two Bi ad-dimers can be described schematically in two
simple steps, as shown in Fig.~\ref{fig:haikucartoon}(b)-(d).  The
first step is to rotate the second and third layer atoms underneath
the Bi ad-dimers, so that the Si dimer atoms sink down into the
surface. This creates the 5- and 7-membered rings also seen in the
As/Ge step structure. The central four atoms are then removed, and the
two halves of the structure bonded together, thus creating the Haiku
structure. (N.B. This is not proposed as the formation mechanism, but
is simply given as a means of understanding the substructure.)  A
linescan of the nanoline has been matched to the Haiku model, using
the background Si dimers as reference marks\cite{Owen2002b}. The
positions of the Bi dimers match up extremely well to the peaks in the
linescan, giving strong confirmation of this model.  Other features of
experimental linescans also agree well with simulated linescans for
the Haiku model\cite{Owen2005b,Miwa2005}.  DFT calculations of this
new structure\cite{Owen2002b} found that it was considerably more
stable than the Miki model, which has been confirmed by subsequent
modelling\cite{Wang2003,MacLeod2005,Miwa2005,Miwa2005b}.  Although the
Haiku structure shares several structural features with previous
models---the top of the structure has two Bi dimers joined by rebonded
second-layer Si atoms, as in the Miki model, while on the outside of
the Bi dimers, there are more rebonded Si atoms, just as in the Naitoh
model---the difference lies in the Si substructure; this mixture of 5
and 7-membered rings extends down many layers.  Consideration of the
core of the nanoline reveals that it is a small triangular section of
hexagonal silicon, embedded in the diamond cubic silicon substrate, 
rather like the ``endotaxial'' RE nanowires mentioned in 
Section~\ref{sec:formation-structure}. The \{111\} planes that delineate the 
core are shown in red in Fig.~\ref{fig:haikucore}.   The
hexagonal core of the nanoline does not exist as a bulk structure for
silicon; it is particular to this nanoline structure. In this respect, the 
Bi nanoline is quite unlike the epitaxial rare-earth silicide family, which are 
essentially one bulk crystal grown epitaxially onto another bulk crystal, taking their
shape from the highly anisotropic mismatch between the two. This core
structure is responsible for many of the properties of the Bi
nanoline. It will be difficult or impossible to kink, resulting in the
extremely straight nanoline observed. As will be shown below in
Section~\ref{sec:DEZ-sub}, the tensile strain field results in the
repulsion of other nanolines, step edges and missing dimer defects. As
a result, the nanolines will not grow sideways, will not grow
together, and will not coarsen into larger islands, unlike the
silicide wires. The electronic structure of the Haiku model has been
studied in some detail\cite{Owen2003,Owen2004,MacLeod2005,Miwa2005b},
and again the nanoline becomes dark in low-bias images --- this
effect is a property of the Bi dimers rather than that of the
nanoline.  Low-bias STM images reveal other electronic
effects, which provide further confirmation of the Haiku structure,
while ruling out the Miki structure. These features are discussed
below in Section~\ref{sec:electronic-structure}.  

However, despite the
wealth of agreement of the predictions from the Haiku model with
experimental evidence, there has been no direct confirmation of the
Haiku structure.  The Bi nanolines are destroyed by exposure to air,
while burial of the nanolines in a capping layer also changes their
atomic structure\cite{Sakata2005,Yagi2005} (though if done carefully
the 1D character and dimerisation of the Bi can be preserved), 
so that cross-sectional TEM studies\cite{Matsuhata2004} have not provided 
definitive confirmation of the core structure, and a recent X-ray Standing
Wave(XSW) study\cite{Saito2003} which relied upon capping with
amorphous silicon and whose results cast doubt on the Haiku model,
cannot be regarded as a counterindication.  The difficulties of burial
of the Bi nanoline are discussed in Section~\ref{sec:burial}.  We note
also that a recent suggestion\cite{Miwa2005b}, that the height of the
Bi dimers in the Haiku model is too far above the plane of the surface
Si dimers based on XPD\cite{Shimomura2000}, is not a significant piece
of evidence against the Haiku model: the XPD data was \emph{fitted} to
the Miki model, thus naturally giving the wrong data for the Haiku.
If a fit of the data to the Haiku model were to give height data which
contradicted the atomistic modelling, then the model would have to be
revisited.

\subsubsection{Miki model revisited}
\label{sec:miki-model-revisited}
\begin{figure}
\includegraphics[width=\columnwidth]{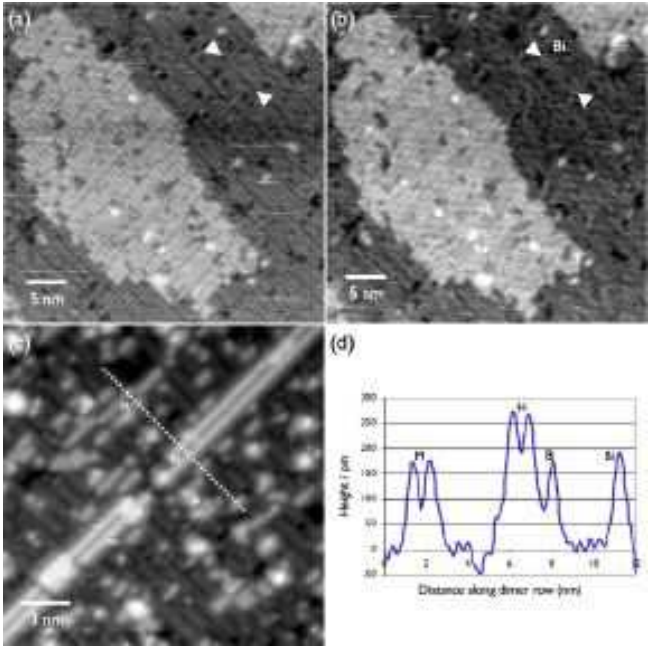}
\caption{STM images of Si(001) surface after deposition of Bi for 70s
  at 400~$^{\circ}$C, followed by 10 min. annealing. (a):
  -2.2~V,0.3~nA. There are many straight missing dimer trenches. (b):
  +2~V, 0.3~nA. Brightening around the trenches in empty-states images
  indicates that they are decorated by Bi dimers, i.e. the Miki
  structure. (c): A Bi-decorated trench and Bi nanoline on a
  H-terminated surface. A linescan along the dotted line (d) reveals
  that the Bi-decorated trenches are significantly lower than the
  nanoline.  (a) \& (b) are reprinted from Appl. Surf. Sci. 142, Naitoh et al.
  Bismuth-induced surface structure of Si (100) studied by scanning
  tunneling microscopy, p.38, (1999), with permission from Elsevier.\cite{Naitoh1999}}
\label{fig:NaitohMiki}
\end{figure}

Despite the apparent success of the Haiku model, the Miki model has
continued to attract a large amount of
attention\cite{Miwa2002a,Miwa2002b,Srivastava2004,Miwa2005,Miwa2005b}
as a possible candidate for the nanoline structure. While in fact its
incorrect registry with the Si substrate unambiguously rules it out as
a candidate nanoline structure\cite{Naitoh2000,Owen2002b,Owen2005b},
it is an energetically favourable structure, and might be expected to
be present on the Bi-covered Si(001) surface.  With the benefit of
hindsight, it can be seen that the Miki structure was indeed observed
in early STM images of the Bi-rich surface.  Fig.~\ref{fig:NaitohMiki}
shows that annealing a Bi-rich surface at 400$^{\circ}$C produces 
a large density of quite straight dimer vacancy (DV) trenches. In the 
filled-states image, Fig.~\ref{fig:NaitohMiki}(a), these appear to be 
composed of clean silicon. However. in the empty-states image,
Fig.~\ref{fig:NaitohMiki}(b), there is contrast between the
dimers adjacent to the missing dimer trenches and the rest of the
surface. This contrast does not occur on the clean surface at this
bias voltage, and so it can be inferred that the trenches are
decorated by Bi dimers, i.e. the Miki structure has been formed. As with the 
Ge-induced DV trenches mentioned in Sec.~\ref{sec:other-nanowires}, 
there is cooperative strain relief between a Bi dimer and a 1DV. Bi dimer/DV
structures have the lowest energy/Bi dimer of any structures except
the Bi nanolines. There is a kinking energy for these
(2$\times$n) trenches of around 0.1eV\cite{Owen2005b}. Hence the
formation of a semi-regular array of Bi dimers reduces the energy further.

Further proof of the Bi decoration of 1DV trenches has been provided
by a more recent experiment, in which a Bi-rich surface was quenched
from 550~$^{\circ}$C at an early stage of the annealing
process\cite{Owen2005b}. Here atomic hydrogen was adsorbed after
cooling to enhance the contrast between Bi and Si significantly, as
the H will adsorb easily to the Si dimers, but does not adsorb on
Bi\cite{Naitoh2000,Owen2002a,Owen2005b}, as can be seen in
Fig.~\ref{fig:registry}.  It was found that the H termination
increased the relative \emph{apparent} height of Bi-related features
by ca.150~pm, or 1.5\AA. An example image is shown in
Fig.~\ref{fig:NaitohMiki}(c).  A linescan across both a Bi-decorated
trench and a section of nanoline shows that the Bi-decorated trench is
quite similar to the nanoline, but is lower and has a deeper
depression between the two Bi dimers.  STM linescans of the feature
decorating DV trenches and the Bi nanolines \cite{Owen2005b} compare
very well to simulated scans of the Miki model and Haiku model
respectively\cite{Miwa2005}, providing strong support for both these
identifications.  Furthermore, in this way, it has been shown that the
Miki structure and the Bi nanoline co-exist, demonstrating once again
that the Miki structure is not the nanoline, as has been recently
suggested\cite{Miwa2005b}, but it is kinetically stable, even at high
temperatures, due to the large activation barrier necessary to form
the Haiku structure.

\subsection{Formation and Properties of Bi nanolines}
\label{sec:form-bi-nanol}
 
\begin{figure}
\includegraphics[width=\columnwidth]{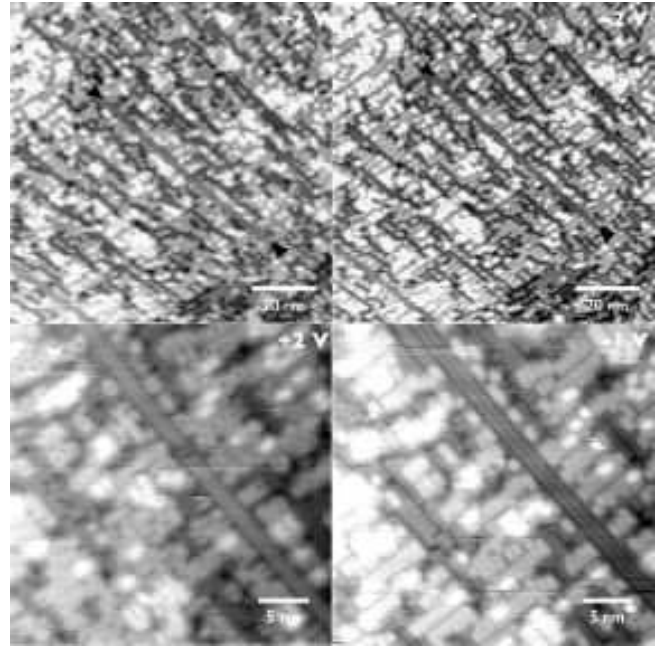}
\caption{STM images of silicon surface onto which 3 ML of Bi has been
  deposited at 360$^{\circ}$C, and then annealed for 60 mins at
  497$^{\circ}$C, resulting in Bi nanolines surrounded by a Bi
  overlayer, which forms a (2$\times$n) reconstruction, with n=4 or 5,
  to accomodate its large strain.  Left-hand images are empty-states,
  right-hand images are filled-states. The (2$\times$n):Bi does not
  grow over the Bi nanoline at this temperature.}
\label{fig:Bi2xn}
\end{figure}

While rare-earth silicide wires have quite specific growth recipes in
order that the nanowires do not coarsen, the Bi nanolines are robust 
to a wide variation in the growth conditions. The essence of Bi nanoline 
formation is that it is a competition between deposition and evaporation.  
A thick layer of Bi on Si(001) is only stable below the Bi bulk evaporation 
temperature (ca. 400$^{\circ}$C).  Above this temperature, an epitaxial layer of 1-2 ML
Bi is stable, and annealing of this surface will result in a surface
with Bi nanolines co-existing with these Bi islands, as shown in
Fig~\ref{fig:Bi2xn}.  The first-layer Bi forms a (2$\times$n)
structure, with n=4 or 5\cite{Naitoh1997,Miki1999a,Bowler2002}, and the
second-layer Bi forms as small groups of dimers, which can be either
parallel or perpendicular to the underlying Bi.

\begin{figure}
\includegraphics[width=\columnwidth]{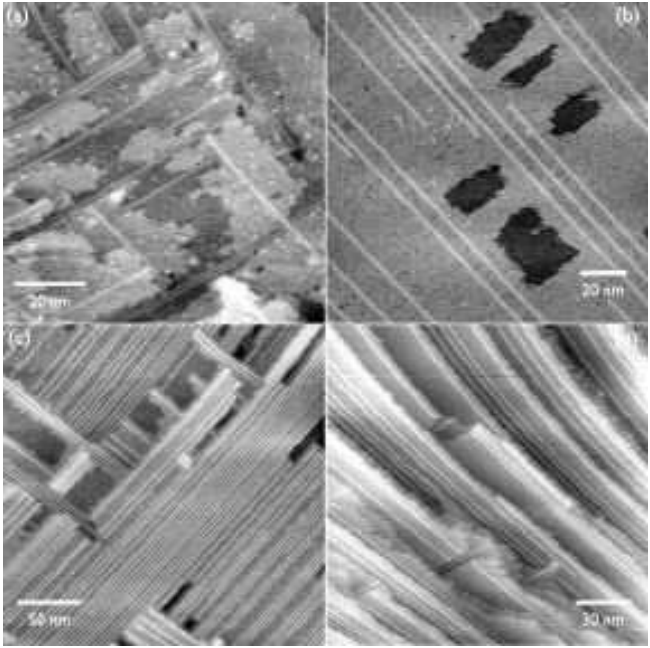}
\caption{Different growth procedures for the formation of Bi
  nanolines, and the resulting surfaces. (a): Deposition of 1 ML Bi at
  a lower temperature, followed by a 20 min. anneal at 520$^{\circ}$C.
  (b): Deposition of 1 ML Bi at around 550$^{\circ}$C, followed by
  annealing at the same temperature. (c): Continuous exposure of 0.1
  ML/min Bi flux for 40 mins at 570$^{\circ}$C. (d): As (c), but on a 0.5$^{\circ}$ miscut
  sample. The growth recipes are summarised schematically in Fig.\ref{fig:recipesplan}. }
\label{fig:recipes}
\end{figure}

The threshold temperature at which Bi dimers on Si(001) will start to
evaporate is around 500$^{\circ}$C, while the maximum temperature at
which any Bi is stable on the Si(001) surface is around
600$^{\circ}$C.  Deposition of Bi in this temperature window, or
deposition of Bi at a lower temperature, followed by an anneal within
this temperature window, will result in a surface comprising Bi
nanolines on an otherwise clean surface, as in Fig.~\ref{fig:recipes}.
However, the details of the growth recipe will determine the surface
morphology at the end of the anneal, as is described below in
Sec.~\ref{sec:recipes-sub}. Of particular technological interest is the growth of
majority-domain surfaces after long anneals. The Si(001) surface has
two equivalent domains with the dimer rows running in orthogonal
directions. On a typical Si(001) surface, these two domains will have
roughly equivalent total areas. Single-domain surfaces form on vicinal
wafers miscut along a $\langle110\rangle$ direction, where the dimer
rows tend to run perpendicular to the step edges. However, by long
anneals of the Bi-rich surface towards the high end of the temperature
window, majority-domain surfaces can be formed, even on flat surfaces.
In this case, extremely long Bi nanolines grow, up to 1$\mu$m, as can be
seen in Fig.~\ref{fig:mcleanimage}.

High-temperature STM observations of the annealing Bi-rich surface
shows a consistent series of surface morphologies. Immediately after
the end of Bi deposition, a relatively rough surface is obtained, with
a high density of small islands, composed of a mixture of Bi and Si,
as shown in Fig.~\ref{fig:recipes}(a).  These will quickly disappear,
but at the same time, nanolines begin to form, either on the terraces,
or on one of the mixed islands.  The nanolines grow out over the step
edges in long, finger-like islands known as promontories, with a Bi
nanoline surrounded by a thin strip of silicon. Examples of the growth
of promontories are seen in Fig.~\ref{fig:promontory}, and more mature
examples of this type of surface are shown in
Figs.~\ref{fig:recipes}\&\ref{fig:mcleanimage}. The nanolines will not
only form long, narrow promontories, but also the reverse, a deep
inlet where a nanoline in one terrance has grown through an up-step
into an upper terrace.  The surface between the Bi nanolines contains
a large density of mobile missing-dimer defects, apparent in STM as
dark, fast-moving streaks.  After further annealing, the missing dimer
trenches begin to disappear, and the Bi nanolines remain on a very
flat, featureless surface. Finally, the Bi nanolines themselves
evaporate. In the following sections, we will give describe the
surface processes which drive this sequence of events, and account for
these observations. The interactions of the Bi nanoline and its
associated strain field with defects and step edges in the terraces,
and with each other, will be discussed and the possible nucleation
mechanisms of this complex structure explored.

\subsubsection{Growth Recipes}
\label{sec:recipes-sub}

\begin{figure}
\includegraphics[width=\columnwidth]{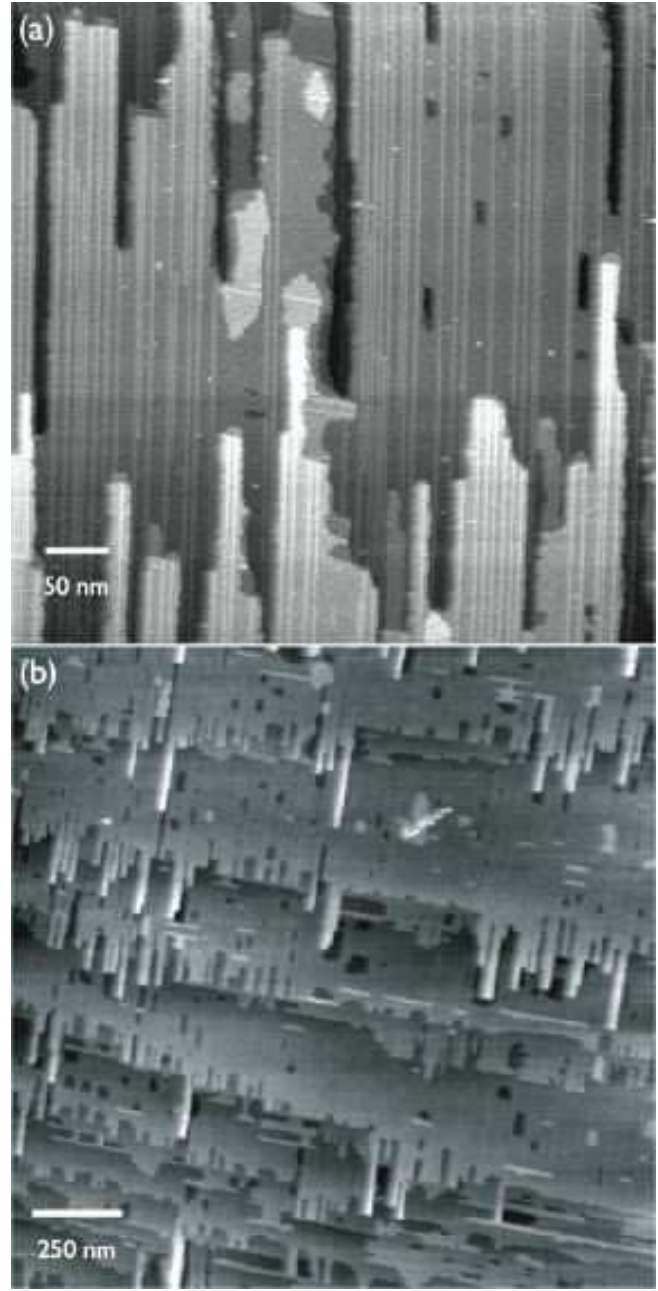}
\caption{(a): An example of a single-domain surface with Bi nanolines
  hundreds of nm in length.  Very long promonotories and inlets are
  present. Reprinted with permission from Ref~\cite{MacLeod2004a}.
  Copyright (2004) by the American Physical Society.
  (b): The larger-scale image shows clearly the single-domain
  nature of the surface, with double-height steps separating
  domains. \textit{Images courtesy of Prof. A.B. McLean, Queen's University, 
  Kingston, Ontario, Canada.} }
\label{fig:mcleanimage}
\end{figure}

\begin{figure}
\includegraphics[width=\columnwidth]{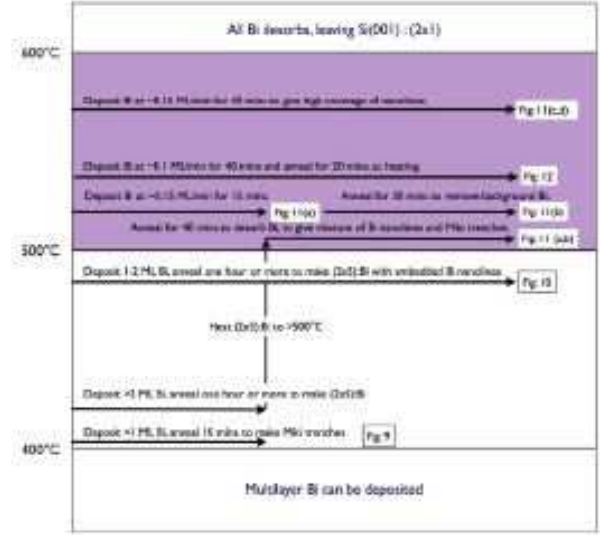}
\caption{Schematic diagram showing the different recipes used to
  produce Bi nanolines.  Below 500$^{\circ}$C, a ML of Bi is stable on
  the surface, and the (2$\times$5):Bi reconstruction results.
  Annealing this surface above 500$^{\circ}$C, or deposition of Bi
  onto the clean surface above this threshold results in Bi nanolines
  on a clean background. High coverages of nanolines are obtained by
  higher-temperature anneals, with longer Bi exposure times.}
\label{fig:recipesplan}
\end{figure}

As shown above, annealing a Bi-rich surface at lower temperatures
results in a surface containing a high density of missing dimer
trenches, decorated with Bi dimers, i.e. the Miki structure. One
feature of the Bi nanoline is the high temperature required to form
it, implying a significant activation barrier to its nucleation.
While the large number of growth parameters, such as
flux, total amount deposited, substrate temperature, miscut angle and azimuth, 
provides a wide range of different possible recipes, as summarised in 
Fig.~\ref{fig:recipesplan}, there are essentially two 
routes to the growth of nanolines\cite{Miki1999a}.  
One growth method for the nanolines is to
deposit a monolayer of Bi below this threshold temperature, and then
anneal the sample until Bi nanolines form. After around 1 hour at
around 500$^{\circ}$C, this will result in a surface with Bi nanolines
and a background Bi-covered surface\cite{Naitoh1997,Miki1999a}.
Continued annealing of this surface will result in the desorption of
the background Bi forming a surface similar to that shown in
Fig.\ref{fig:recipes}(a), and eventually a surface similar to that
shown in Fig.~\ref{fig:recipes}(b) is obtained.  The second growth
preparation route involves the deposition of Bi onto the surface
within the desorption temperature window. Due to the lower surface
coverage of Bi, and the higher temperature, a more ordered surface is
obtained, and fewer, longer, Bi nanolines form. The final surface is
much the same, however, corresponding to that shown in
Fig.\ref{fig:recipes}(b).  A high density of Bi nanolines can be built
up by continuous Bi exposure for a long time, at a temperature in the
upper half of the desorption window. In this regime, the nanolines are
only stable for a few minutes in the absence of a Bi flux and small differences
in stability, such as the ends of a nanoline, will be significant. Shorter nanolines,
perpendicular to the prevailing direction, will tend to dissolve, so that a highly 
ordered surface with long, parallel nanolines is obtained, as in
Fig.\ref{fig:recipes}(c). By use of a vicinal surface, the preference
for one domain can be enhanced further, and a surface close to a
single-domain surface can be obtained, as in Fig.\ref{fig:recipes}(d).
A second method which produces a majority-domain surface, even on a
surface without any intentional miscut, (the miscut angle is
$\langle0.5^{\circ}$) \cite{MacLeod2004a,MacLeod2004b}, produces a
surface as shown in Fig.~\ref{fig:mcleanimage}. In this case, a
substrate temperature around 530$^{\circ}$C was used, 4.5 ML of Bi
were deposited, with an anneal for around 40 mins.  Comparison of 
Fig.\ref{fig:recipes}(d) and Fig.~\ref{fig:mcleanimage} reveals that in the former
case, the nanolines are parallel to the prevailing step edges, while in the latter, 
they are perpendicular. Despite the large difference in morphology, the only 
differences between this recipe and that which produced Fig.\ref{fig:recipes}(c,d) 
are that continuous deposition for an extended period was not used, and the 
heating current used was AC, a strategy which was designed to reduce 
electromigration. Whether this is is the cause of the majority-domain surface 
remains unclear, but such large-scale single-domain surfaces are ideal for 
nanoelectronics applications.

\begin{figure}
\centering
\includegraphics[width=0.9\columnwidth]{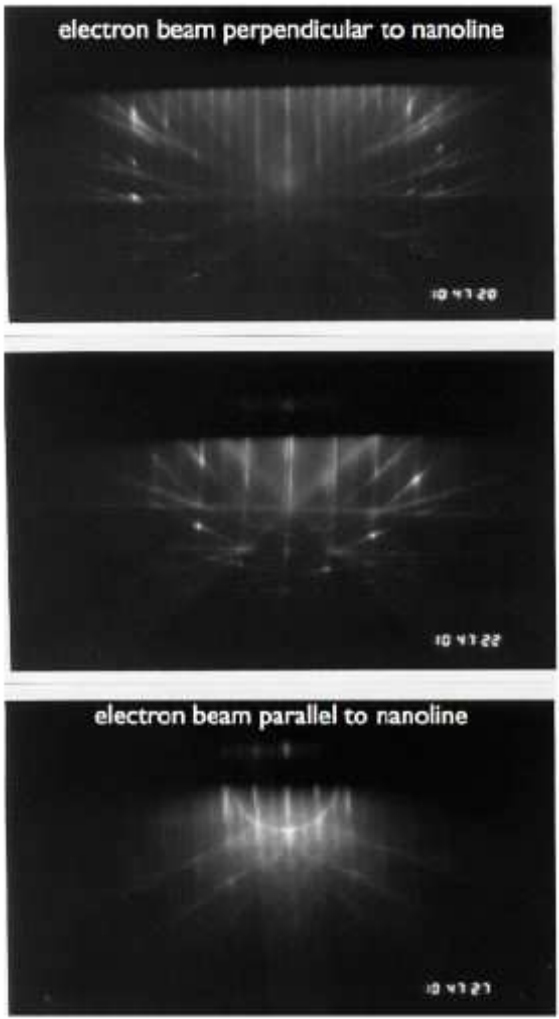}
\caption{A series of RHEED images of a Bi nanoline surface. Along the [1$\bar{1}$0] direction, perpendicular to the nanolines, a series of long streaks is seen, as in (a). Along the  [110] direction, parallel to the nanoline, as in (c), a distinctive ring feature is seen, while at 
intermediate angles (b), the ring opens up into a long arc.}
\label{fig:RHEED}
\end{figure}

While STM can observe only a relatively small part of the surface, the
large-scale order of the surface can be determined using electron
diffraction methods such as RHEED. A series of RHEED patterns at
different azimuthal angles of a single-domain Bi nanoline surface are shown
in Fig.~\ref{fig:RHEED}.  Perpendicular to the nanolines, the RHEED pattern 
comprises vertical lines, while parallel to the nanolines, a ring feature is seen.
The strength of these features demonstrates the
long-range order generated by the nanolines, and means that their
growth and development can be followed in real-time.

\subsubsection{The Bi reservoir}
\label{sec:backgroundBi-sub}

In contrast to the silicide nanowires, which form very rapidly at the
growth temperature and are only stable for about 10~mins. at that
temperature, there is a significant incubation period before the
nucleation of the Bi nanolines, and they will continue to grow for
upwards of an hour after the deposition of Bi has finished. Thus the
Bi required to generate the nanolines must come from a reservoir of Bi
which is present on the surface.  Where is this Bi located, and how
does it affect the growth of the nanolines?

\begin{figure}
\includegraphics[width=\columnwidth]{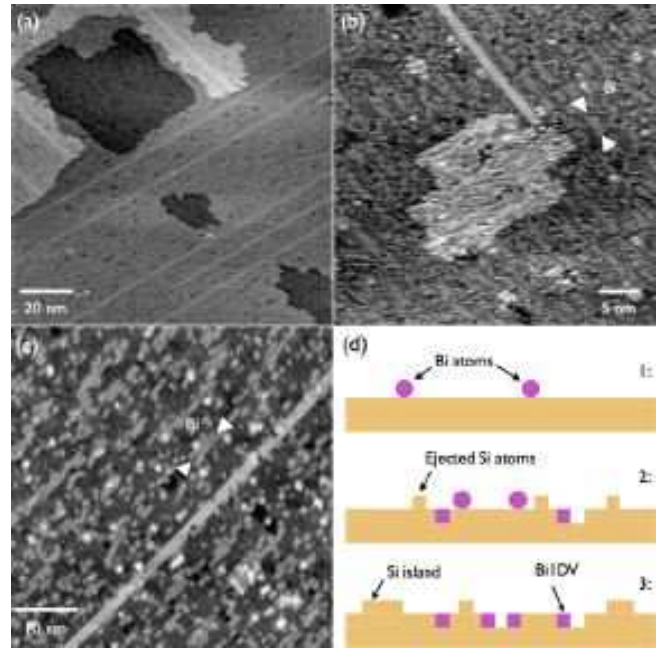}
\caption{(a): Hot STM image of the mature nanoline surface, showing Bi
  nanolines and dark missing dimer defect rows. Any background Bi is
  invisible. (b): Hot STM image of the Bi:Si(001) surface taken at
  about 500$^{\circ}$C, with a sample bias ca. +3.0~V. Grey linear
  features, which run parallel to the Bi nanolines, can be seen in the
  background. An example is marked by white arrows. (c): STM image
  taken at room temperature, with the silicon passivated with hydrogen
  to increase the contrast between the Si and Bi. The grey linear
  feature can now be seen to comprise Bi dimers decorating a
  missing-dimer trench, i.e. the Miki structure. (d): A cartoon of the
  formation of the Bi reservoir. Impinging Bi atoms (1) exchange with
  Si atoms, making Bi surface dimers and Bi!DV complexes (2). The
  ejected Si atoms difffuse and form islands
  (3).\label{fig:reservoir}}
\end{figure}

In images of the surface taken soon after the end of the Bi
deposition, for example, Fig.\ref{fig:recipes}(a), there is often a
high density of small islands on the surface.  These islands have the
same appearance as the rest of the surface; they are not islands of Bi
dimers, as may be seen after deposition at lower temperatures, but
appear to be mixtures of Si and Bi atoms. In general, such small
islands are not stable at these temperatures, as has been seen in hot
STM studies of Si homoepitaxy on
Si(001)\cite{Voigtlander1997,Owen1997c}.  During growth, these small
islands are stabilised by the high flux of Si atoms, but if the flux
is cut off, a process similar to Ostwald Ripening occurs, in which
these islands will decompose, and the material in them will move to
step edges (which may be regarded as a very large island).  The
similarity of the behaviour observed here suggests that these islands
are sustained by a flux of Si and Bi, which is produced during Bi
deposition.  For low coverages of Bi, the lowest-energy sites for Bi
dimers which have substituted for Si dimers in the top layer of the
Si(001) surface---hereafter ``Bi surface dimers''---are Bi surface
dimer/missing-Si-dimer complexes\cite{Owen2005b}, where a Bi dimer
decorates one or both sides of a 1DV, as seen in
Fig.~\ref{fig:NaitohMiki}.  For each adsorbed Bi dimer, therefore, up
to four Si atoms are ejected from the surface layer, producing a
significant transient flux of Si atoms; this flux in turn produces the
observed islands. At the end of Bi deposition, this flux will die
away, the islands will no longer be stable and will decompose, as
occurs between Fig.\ref{fig:promontory}(a)and (b). A cartoon of this
process is shown in Fig.~\ref{fig:reservoir}(d).

Once the small islands have annealed away, the surface morphology is
very flat, and the background between the nucleating Bi nanolines
appears featureless in hot STM, apart from missing dimer trenches.
However, at this stage a large surface density of Bi may be inferred
from the fact that Bi nanolines continue to nucleate and grow on this
surface. The Bi dimers embedded in the surface layer have
approximately the same contrast as the Si dimers\cite{Owen2005b}, and
are therefore invisible at elevated temperature, as in
Fig.~\ref{fig:reservoir}(a). At a large positive bias, however, the
background Bi becomes visible, as is shown in
Fig.~\ref{fig:reservoir}(b). Top-centre of this image is a Bi nanoline.
There is a high density of linear grey features in the background, 
one of which is marked with a pair of white arrows.  Detail
of these features cannot be discerned at this temperature, although
the background Bi is noticeably lower than the nanoline. After quenching
the sample to room temperature, and exposure to hydrogen, we can judge
the true coverage and distribution of Bi during the annealing process,
as displayed in Fig.~\ref{fig:reservoir}(c).
There is a high density of Bi surface dimers, and also many Bi dimers
are adsorbed to one or both sides of a missing dimer trench, forming
the Bi1DV or Miki structures, as shown in Fig.~\ref{fig:NaitohMiki}.
Thus the Bi reservoir consists not of mobile ad-dimers, but mostly of
Bi surface dimers, usually decorating the missing-dimer trenches to
form the Miki structure.

\begin{figure}
\includegraphics[width=\columnwidth]{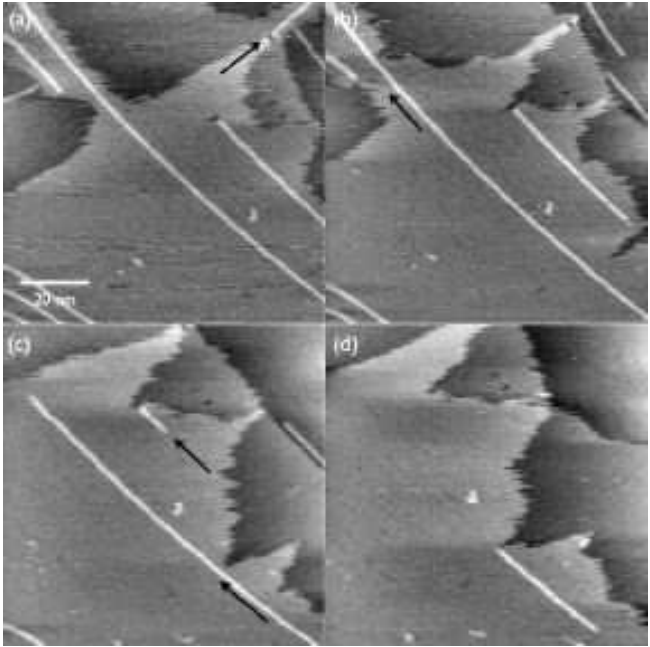}
\caption{A sequence of hot STM images, showing the breakup and
  evaporation of the BI nanolines at high temperatures. Arrows
  indicate the major areas of change from one image to the next.}
\label{fig:linebreakup}
\end{figure}

During the high-temperature anneal, the surface Bi reservoir is
continuously depleted, with some Bi going to form nanolines, and the
remainder evaporating back into vacuum. The activation barrier for
evaporation is different for different surface adsorption sites. A
RHEED study\cite{Miki1999a} found that the activation barrier for
evaporation of Bi from the nanoline was 0.25 eV higher than the
barrier for evaporation of Bi from a (2$\times$n):Bi overlayer. Hence
Bi is mostly being lost from the background reservoir. However, it
might be expected that missing dimers in the nanolines would be
observed occasionally. In fact, this is not the case, and those
nanolines which are observed at high temperatures always appear
perfect.  It is likely that, due to the large defect energy of the
Haiku structure (0.66~eV from DFT calculations\cite{Wang2005}), any
Bi which evaporates from the nanoline is quickly replaced from the Bi
reservoir. Thus at high temperatures, where Bi can evaporate from the
nanolines, their stability is dependent upon a sufficient quantity of
Bi in the reservoir to replace missing dimers. At the later stage of
an anneal, no new Bi nanolines nucleate, and those which are present
cease to grow. This is an indication that the reservoir of Bi is
approaching exhaustion.  Finally, when the reservoir is exhausted,
evaporation from the Bi nanolines will take place, and they will
disappear within a few scans, or sometimes within the time of one
scan. A sequence showing the evaporation of nanolines is shown in
Fig.\ref{fig:linebreakup}. In general, the nanolines evaporate from
the ends, rather than breaking up into pieces, which strongly suggests
that the ends of nanolines are less stable than the middle of the
nanoline, and that the energetic barrier to break the nanoline in the
middle and create two ends is prohibitive, even at these temperatures.

\subsubsection{Promontory Growth}
\label{sec:promontory-sub}

\begin{figure}
\includegraphics[width=\columnwidth]{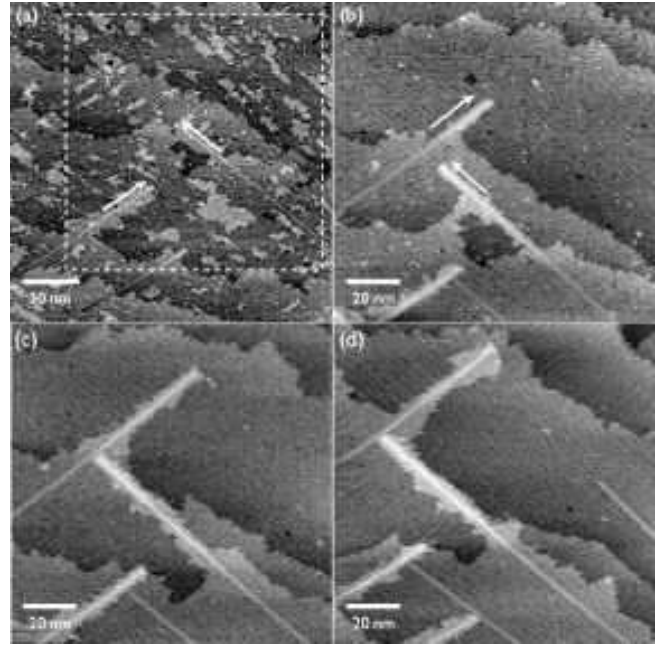}
\caption{A sequence of high-temperature STM images showing the growth
  of a promontory.  Between the first and second images, the high
  density of islands has annealed away, and two nanolines have
  nucleated. The growth of nanolines in the succeeding image is
  indicated by the white arrows. The promontory is a growth effect,
  and not caused by etching away of the surrounding Si.}
\label{fig:promontory}
\end{figure}

The most notable feature of the mature nanoline surface as seen in
Fig.~\ref{fig:recipes}(c) and (d) as well as
Fig.~\ref{fig:mcleanimage} is known as the promontory. Long, narrow
islands of Si surrounding Bi nanolines grow out across lower terraces.
A series of images showing the early stages of Bi nanoline growth are
shown in Fig.~\ref{fig:promontory}.  In (a), a short nanoline has
grown out over a lower terrace and the joining of the promontory and
the pre-existing step edge has formed a hole in the upper terrace, as
is marked in (b). The growth of this promontory is blocked by the
growth of a second nanoline, which has nucleated in the time interval
between (a) and (b), and is growing out across the next lower terrace.
Comparison of (b),(c) and (d) makes it it very clear that the
promontories are the result of growth out over a lower step edge.  By
(d), a common pattern of zigzagging Bi nanolines which have blocked
each other's growth is developing.

The mechanism behind this promontory growth has not been determined;
in an early paper, it was suggested that the promontories form by
etching of Si around the Bi nanolines\cite{Miki1999a}.  However, in
Fig.~\ref{fig:promontory}, there is no sign of etching taking place;
this is a pure growth phenomenon. It is likely that the growth of the
nanoline is related to the high mobility of step edges at
500$^{\circ}$C.  Nanolines will always continue to grow until they
reach an obstacle. Thus nanolines will stop at step edges. However,
the position of the step edge at this temperature is not constant; it
fluctuates back and forth\cite{Zandvliet1993}, as material moves
randomly along it. This motion may be inferred from the jagged,
streaky step edges in these high-temperature images, as discussed in
Section~\ref{sec:hot-stm}.  Therefore, each time the step edge moves
forwards, the nanoline can extend to the new step position, and then
when the step edge tries to move back again, its position is pinned by
the presence of the nanoline.  With each forward fluctuation
therefore, the nanoline will grow forward, and thus a promontory will
gradually form.  A similar process could explain the retreat of an up
step ahead of a growing nanoline.
The notable feature of the promontory is that they have a fixed
minimum width. At least 3-4~nm of Si is maintained on either side of
the Bi nanoline. This fixed width is an example of the phenomenon
known as the Defect Exclusion Zone, or "DEZ", which is discussed in
Section~\ref{sec:DEZ-sub}.

\subsubsection{The role of strain in surface interactions}
\label{sec:DEZ-sub}

Since the nanoline has a strain field associated with it, it is likely
to exhibit quite strong interactions with other surface features which
exert a localised strain on the surface. Such features might be
adsorbed Bi dimers, missing dimer defects, step edges, and other
nanolines. 

\begin{figure}
\begin{center}
\includegraphics[width=\columnwidth]{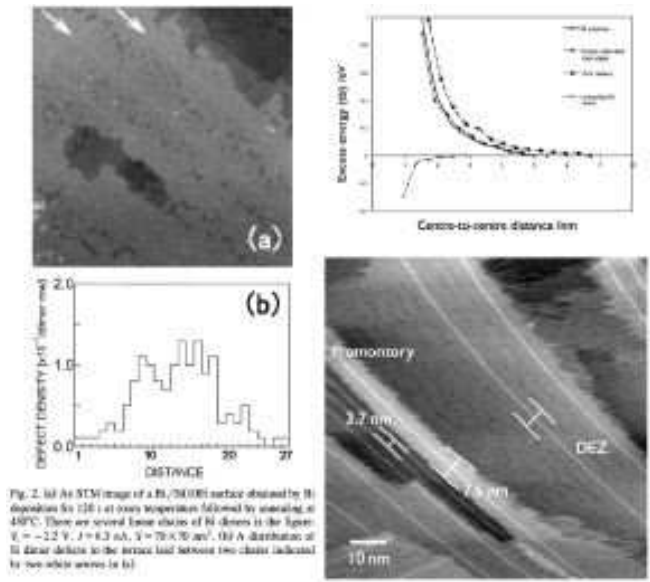}
\caption{Left-hand side.: STM showing parallel Bi nanolines. The
  positions of missing dimer defects in between the two nanoline
  marked with white arrows have been gathered and are shown ini the
  graph below.  (Reprinted from Appl. Surf. Sci. 412, Naitoh et al.
  Bismuth-induced surface structure of Si (100) studied by scanning
  tunneling microscopy, p.38, (1999)\cite{Naitoh1999}, 
  with permission from Elsevier.)
  Right-hand side: Graph of interactions between nanoline and other
  surface features - 1DV defects, step edges, Bi surface dimers, and
  other nanolines. Distances are measured from the centre of the
  nanoline to the centre of the feature.  High-temperature image
  showing Bi nanolines and disordered dark trenches of Si missing
  dimers. Although the defects are very mobile, there is an empty
  space around each nanoline, known as the Defect Exclusion Zone, or
  'DEZ'. The DEZ is about 7 nm across.  This width is the same as the
  width of the promontory. }
\label{fig:DEZnewfig}
\end{center}
\end{figure}

Using tight-binding simulations, where very large cells can be
relaxed, the interaction between a nanoline and several features, such
as 1DV defects and rebonded B-type step edges have been calculated.
These data are plotted in Fig.\ref{fig:DEZnewfig}.  From
elevated-temperature STM images, it is clear that there is a repulsive
interaction between the nanolines and missing dimer defects and step
edges. At high temperature, as in Fig.~\ref{fig:DEZnewfig}, the
surface is covered in black streaks, which indicate the positions of
rapidly moving missing dimer defects. However, either side of the
nanoline, there are no streaks. Likewise, where a nanoline forms a
promontory, the width of this empty area is continued as the width of
the promontory. This characteristic width of defect-free silicon,
around 3-4~nm either side of the nanoline, is known as the ``Defect
Exclusion Zone'' or DEZ \cite{Owen2004}. For a large separation, the
total energy of a nanoline and a defect is the same as for the two
structures independently. However, as the separation decreases, the
total energy increases, so that at a distance of about 3.5 nm, there
is a repulsive interaction between the nanolines of ca. 0.1 eV for
both step edges and 1DVs, and at smaller distances, this
energy increases rapidly. Statistics about the
position of 1DVs in the space between two nanolines
have been gathered, as shown in Fig.~\ref{fig:DEZnewfig}, taken from
Ref.~\cite{Naitoh1999}. They show that on average, most defects occur
at least 3.2 nm away from the nanoline, which is in agreement with the 
calculated repulsion. 

For a pair of nanolines, there is a similar interaction, and the 0.1
eV threshold is also reached when the centres of the two nanolines are
about 3.5 nm apart, or in other words, the gap between the nanolines
is the width of one nanoline.  Clusters of nanolines with this
approximate spacing are often seen, as in Fig.~\ref{fig:DEZnewfig}.
For nanolines adjacent to each other, the excess energy is about 1 eV
per unit cell. Despite this, nanolines are sometimes observed growing
next to each other, as in Fig.~\ref{fig:recipes}(c). The reason for this is that
unlike defects and step edges, which are mobile, a nanoline cannot move sideways.
Hence when two nanolines which have nucleated in different places grow
past each other, they cannot move sideways to reduce their interaction
energy. The presence of nanolines in close proximity does indicate that 
this interaction energy is not so high as to present an
insuperable barrier to the growth of two nanolines past each other.
For Bi surface dimers, which have a compressive stress field,
decoration of missing dimer trenches is a natural way in which to
relieve their stress. The gain in energy by this process is about 0.45
eV/Bi dimer in tightbinding calculations\cite{Owen2005b}.  The
interaction with the nanoline is also attractive.  In images where
there is a large surface density of Bi dimers, they are often situated
adjacent to the nanolines.  This position is found to be 0.28 eV
better in energy\cite{Owen2005b} than elsewhere.  This suggests that
the nanolines are generally more stable in a surface with a
significant amount of compressive stress, such as one which is covered
in Bi. Indeed the energy of a nanoline in a Bi-terminated surface is
0.8 eV better than on a Si-terminated surface.

\subsubsection{Nucleation}
\label{sec:nucleation-sub}

\begin{figure}
\begin{center}
\includegraphics[width=\columnwidth]{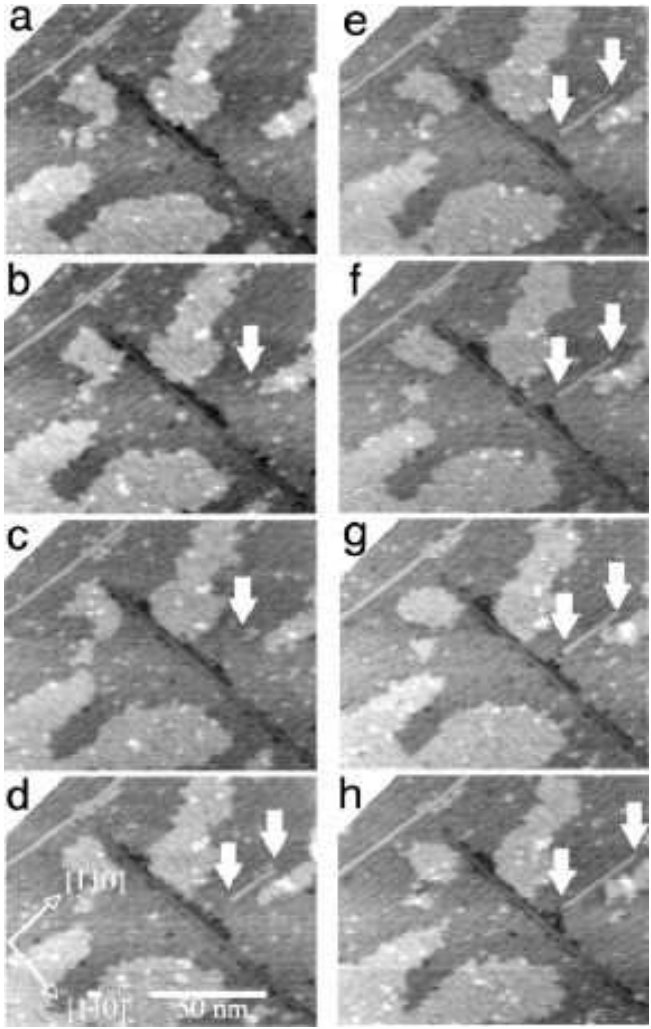}
\caption{Example of observation of the nucleation of the Bi line at
  497$^\circ$C. A series of 81~nm STM images. During the interval from
  (b) to (c) of 14 min., the Bi line on the bottom extended to
  separate the big Bi island and to make an inlet as marked by a white
  arrow.  During the same period, a new Bi line appeared as marked by
  a black arrow.  (d) to (e) shows more details of observation
  corresponding this latter event, in the area marked by a white box
  in (c). The time between images was 18 sec.  In figure (e) a small
  Bi island is situated at the place marked by a black arrow, in (f)
  the small Bi island changed into a short Bi line, and between (f)
  and (g) the new line extends in both directions towards the island
  and the inlet. It is noted that the side edges of the inlet, and the
  shapes of all the islands change image by image. Reprinted from
  Surf. Sci. 421, Miki et al.  Bi-induced structures on Si(001),
  p.397, (1999), with permission from Elsevier.  }
\label{fig:nucleationSTM}
\end{center}
\end{figure}

Despite the success of the Haiku structure in explaining all of the
above surface phenomena, its identification raises as many questions
as it answers. How would such a complex structure form? How is it
terminated? If termination is energetically unfavourable, what would a
nucleus structure look like?

Experimentally, the atomistic nucleation process remains a mystery.
There is an incubation time during the annealing process before
nanolines start to nucleate, which suggests that there is a
significant barrier to formation of these structures. On many
occasions, the appearance of Bi nanolines has been captured during STM
observations, such as in Fig.~\ref{fig:nucleationSTM}. There is no
apparent precursor state; like Athena from Zeus' brow, a short nanoline 
springs fully-formed from an empty patch of the surface. The
nanolines then grow extremely rapidly.

\begin{figure}
\begin{center}
\includegraphics[width=0.9\columnwidth]{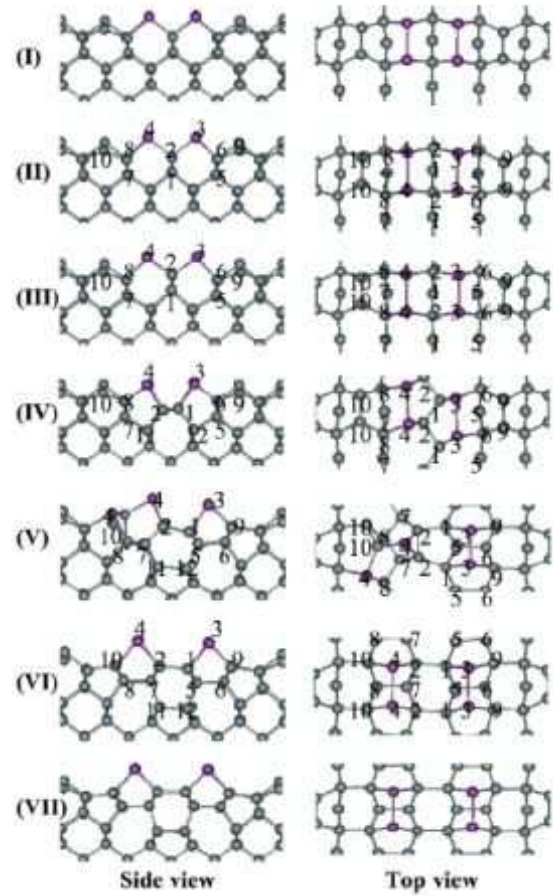}
\caption{A proposed mechanism for the nucleation of a Bi nanoline. A
  Bi ad-dimer fills a Bi1DV trench, to form a pair of adjacent Bi
  dimers. In order to relieve its strain, the atoms between the Bi
  dimers twist sideways to form a zigzag structure.  Significant
  reconstruction takes place between (iv) and (v), resulting in the
  Haiku structure. The barrier for this process is less than 1.5~eV.
  Reprinted with permission from Ref.~\protect\cite{Wang2005}.
  Copyright (2005) by the American Physical Society.}
\label{fig:nucleation}
\end{center}
\end{figure}

Thus far, there has only been one proposed mechanism for the
nucleation of a Bi nanoline\cite{Wang2005}.  It has been suggested
that the initial step is for Bi ad-dimers to fall into the missing
dimer trenches on the hot Si(001) surface, forming a row of pairs of
Bi dimers.  Such a structure would result in the increase in local
compressive stress, which may provide a driving force for nanoline
nucleation.  We demonstrated earlier (Fig.\ref{fig:haikucartoon}) that
the 7-5-7 structural motif may be achieved by a simple bond rotation
process.  Using this motif, the proposed mechanism suggests that the
central four atoms between the two Bi dimers in the proposed mechanism
twist, so that a zigzag core structure is formed.  From this the Haiku
structure evolves, with a maximum barrier in the calculation of less than
1.5~eV.  The reaction pathway and associated energy diagram are shown
in Fig.~\ref{fig:nucleation}.  This mechanism has a number of
interesting features, but also a major limitation. The calculation
takes place in a periodically repeated unit cell one dimer row wide,
and is therefore considering the nucleation of an infinite nanoline,
\emph{without} ends. It is suggested that a long nucleus forms by the
filling of several unit cells of a missing dimer trench, but this is
an unstable situation\cite{Owen2005b}.  Moreover, in our STM
observations, we observe the appearance of very short nanolines, which
then grow. The termination of these finite nuclei \emph{must}
therefore be considered.  STM observations of nanoline evaporation
show that nanolines do not break up in the middle, instead they
evaporate from the ends. From this, it may be inferred that the energy
of termination is quite high.  While there has been no systematic
study of termination, we have performed calculations on many and
varied structures, all of which have been found to have a high energy.
In tightbinding modelling, the energies were at least 1~eV/end , while
those terminations which we have considered for the proposed zigzag
nucleus structure have much higher energies, around +1.6~eV/end,
giving a total nucleus energy of over +4~eV.  This suggests both that
understanding the termination of the nanolines is an important area
for future research, and that nucleation structures should be
considered in 2D, rather than in 1D (which may not be feasible with a
technique more accurate than tightbinding).  Given that the prevalent
structures on the Bi-rich surface are Bi surface dimer/1DV complexes,
alternative nucleation mechanisms which use these structures as a
basis may be more fruitful and are being explored.

\subsection{Electronic Structure}
\label{sec:electronic-structure}

The knowledge of the electronic structure of the Bi nanoline has
stemmed from two sources: variable-voltage
STM\cite{Miki1999a,Owen2003,MacLeod2005} and DFT calculations of the
electronic
structure\cite{Srivastava2004,Owen2002b,Bowler2000,Miwa2002a,Bowler2002,Miwa2002b,Owen2003,MacLeod2005,Miwa2005b}
(generally considered in terms of the local density of states (LDOS),
or the Tersoff-Hamann approach to STM, which uses the LDOS).
Knowledge of the atomic structure of the Bi nanoline allows detailed
calculations of its electronic structure to be made, which can then be
compared to the observations made in STM.  While the majority of the
electronic contrast is expected to be dominated by the Bi dimers in
the nanoline, other details of the LDOS have allowed a stronger
identification of the Haiku structure with the Bi nanoline to be made,
with the other possible candidate structures ruled out after
comparison with the STM data.

\begin{figure} 
\includegraphics[width=\columnwidth]{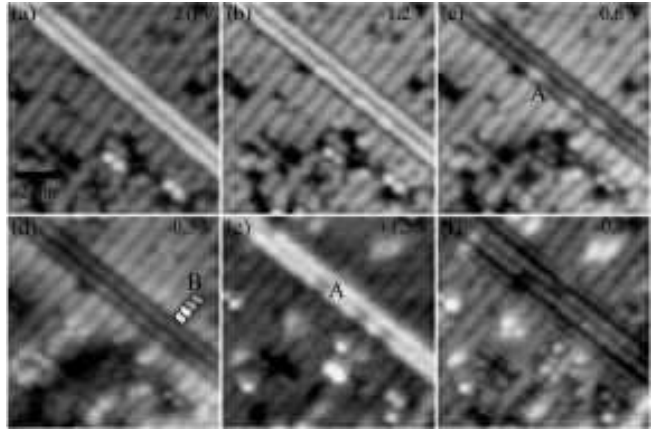} 
\caption{10\,nm $\times$ 10\,nm images of a Bi nanoline. The sample
  bias voltages used are -2.0\,V, -1.2\,V, -0.6\,V, -0.3\,V, +1.5\,V
  and -0.8\,V, in (a-f) respectively. As the sample bias is reduced,
  between (a) and (d), the nanoline changes contrast from light to
  dark relative to the surrounding Si(001). Over this range, some of
  the dimers in the nanoline (marked A in (c) and (e)) exhibit a
  different voltage contrast. At very low biases, around -0.3\,V, an
  enhancement of the dimers around the nanoline, similar to that seen
  around a missing dimer defect in clean
  Si(001)\protect\cite{Owen1995}, is seen.  This is visible (and
  marked schematically as B) in (d). In (f), the resolution is
  sufficient to see that the corrugation of the Si dimers closest to
  the nanoline is increased, (marked schematically by the dotted black
  lines) suggesting a greater separation, and hence tensile strain.
  Reprinted from Surf. Sci. Lett. 527, Owen et al. Interaction between
  strain and electronic structure in the Bi nanoline, p.177, (2003),
  with permission from Elsevier.\label{fig:varbias}}
\end{figure} 
 
We begin by summarizing the data available from STM, which is
illustrated in Fig.~\ref{fig:varbias}:

\begin{enumerate}
\item At high biases, the Bi nanolines appear bright compared to the
  surrounding Si(001) surface\cite{MacLeod2004a,MacLeod2004b,Miki1999a,Miki1999b,Naitoh1999,Naitoh2000,Owen2005b,Owen2002b,Owen2003,MacLeod2005,Miwa2005,Naitoh1997,Naitoh2001}
\item At low biases, the Bi nanolines become dark compared to the
  surrounding Si(001) surface\cite{Miki1999a,Miki1999b,Owen2005b,Owen2003,MacLeod2005}
\item At low biases, the Si dimers neighbouring the Bi nanoline show
  enhancement, or brightening\cite{Owen2003} similar to that seen
  around missing dimer defects on Si(001)\cite{Owen1995,Schofield2004}.
\end{enumerate}

These facts suggest various things: first, that the Bi dimers are
physically higher than the surrounding Si(001) surface dimers, as at
high biases the contrast in STM is determined primarily by geometrical
structure; second, that the \emph{local} electronic structure of the
Bi dimers will show states further from the Fermi level than the
surrounding Si dimers\cite{Owen2003,MacLeod2005,Miwa2005b}, as this
determines the STM current at low biases; finally, that the Bi
nanoline structure must be under tensile stress, which causes the Si
dimers near the nanoline to be pulled together, away from their
relaxed surface structure, which is the origin of enhancement at low
biases\cite{Owen1995,Schofield2004}.

\subsubsection{Electronic Properties of the Bi dimer}
\label{sec:properties-bi-dimer}

\begin{figure}
\centering
\includegraphics[width=0.9\columnwidth]{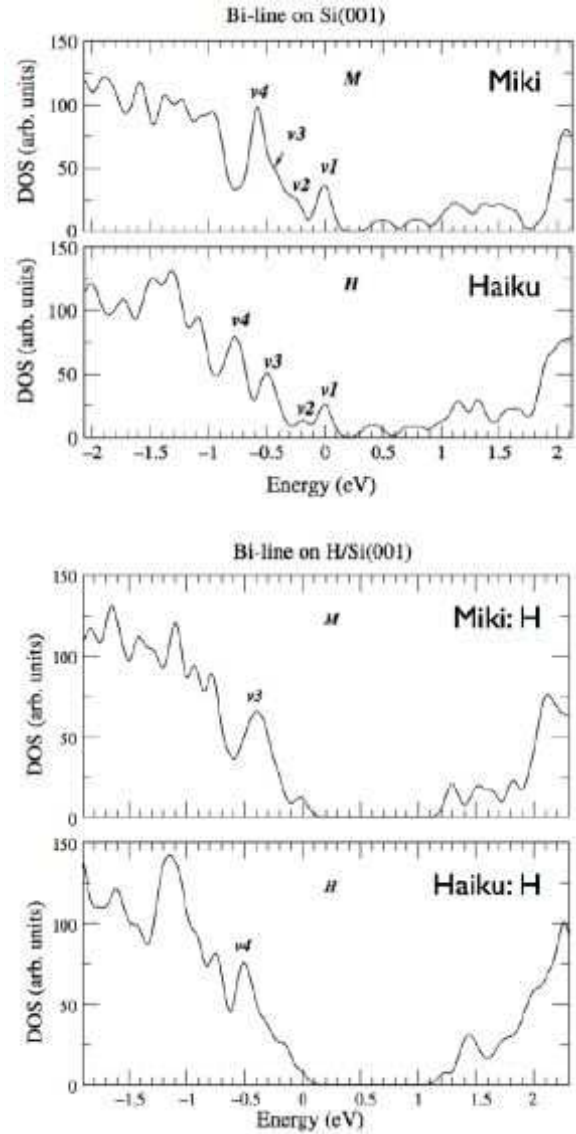}
\caption{The calculated DOS of the Miki and Haiki structures on a
  clean surface (top) and a H-terminated surface (bottom) are shown.
  The marked states v1 to v4 are described in the text.  Reprinted with
  permission from Ref~\cite{Miwa2005b}. Copyright (2005) IOP
  Publishing.}
  \label{fig:MiwaDOS}
\end{figure}

Examination of the detailed electronic structure of the Miki and
Naitoh models\cite{Miwa2002a,Miwa2002b,Srivastava2004} and the Haiku
structure\cite{Owen2003,Owen2004,MacLeod2005} all show that the Bi
dimers become dark at low bias voltages by contrast to the surrounding
silicon dimers.  The calculated DOS for the Miki and Haiku structures
are shown in Figure~\ref{fig:MiwaDOS}\cite{Miwa2005b}.  The $v1$ peak
comes from the up atoms of the substrate Si dimers, as does the $v2$,
while the $v3$ peak comes from both the Bi and Si dimer bonds and $v4$
from the Bi atoms in the line models.  When the substrate dimers are
passivated with H, the peaks $v1$ and $v2$ almost disappear leading to
peaks in the gap which arise from the Bi dimers ($v3$ and $v4$).  This
data confirms earlier modelling and bias-dependent STM
measurements\cite{Owen2003} discussed below.

Consideration of the electronic structure of the Bi dimer reveals that
it is rather unusual.  In its native form, Bi forms puckered hexagonal
sheets, similar to the structure taken by As.  Full-potential LAPW
modelling of Bi in a bulk diamond cell (thus forcing it to take up
tetrahedral bonding) showed that there is very little hybridisation
between the s- and p-orbitals\cite{Bowler2000}.  Thus we expect Bi to
tend towards p$^3$ bonding with a filled s-orbital (as a lone pair)
and to be energetically most favourable with 90$^\circ$ bond angles.
This explains to some extent the Haiku structure, where bond angles
are almost 90$^\circ$.  Projection of the local densities of states
calculated using pseudopotential-based DFT
calculations\cite{MacLeod2005} suggests that the Bi dimer in the Haiku
structure has hybridisation between s and p$_{\mathrm{z}}$, with
p$_{\mathrm{x}}$ and p$_{\mathrm{y}}$ separated off.  The FLAPW
calculations showed that the relativistic effects of the full core
increased the s-p splitting in the Bi atom; while relativistic effects
can be incorporated in pseudopotentials, the Haiku structure might be
more stable, and show bonding closer to p$^3$, if these effects were
taken fully into account\footnote{We note that a full potential
  calculation on a unit cell large enough to hold the haiku structure
  is probably computationally prohibitive}.

\begin{figure}[htbp]
  \includegraphics[width=\columnwidth]{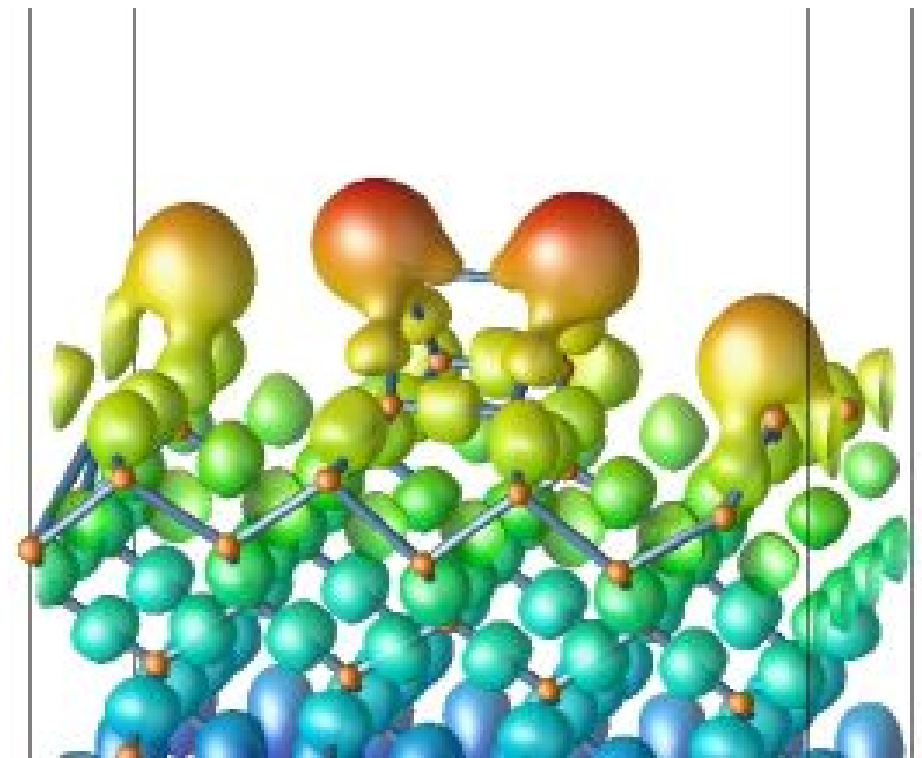}
  \includegraphics[width=\columnwidth]{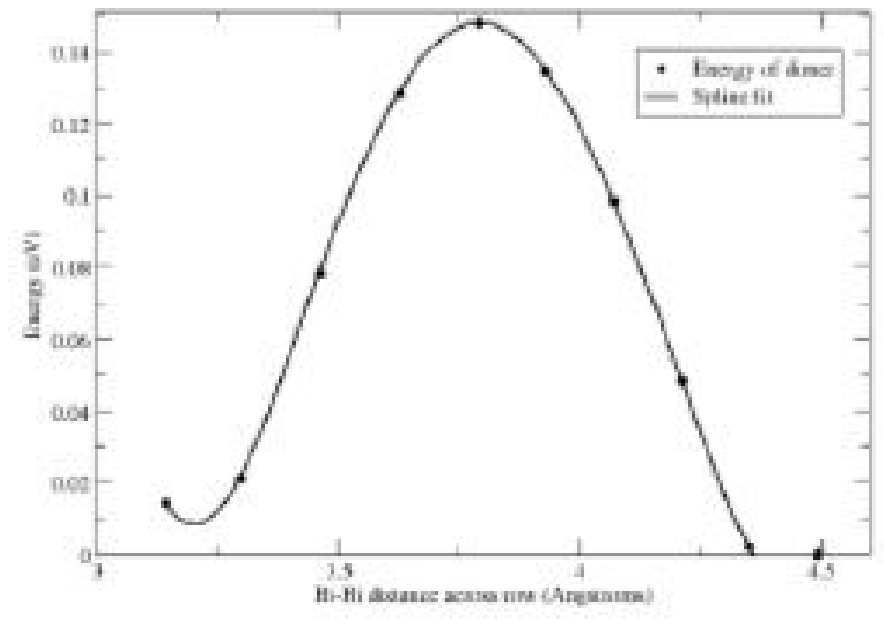}
  \caption{(a) The electron localisation function (ELF) of a Bi dimer
    on Si(001)\cite{Bowler2005} for a value of 0.8.  The ELF indicates
    the localisation of pairs of electrons, indicating strong bonding
    and lone pairs.  The figure shows the weakness of the Bi-Bi bond,
    and the diffuseness of the Bi lone pair.  See text for full
    discussion. (b) Energy barrier for Bi dimer to go from over dimer
    row to over trench between dimer rows.  The dotted line is a
    spline fit intended as a guide for the eye.\label{fig:BiDimerELF}}
\end{figure}

The Bi-Bi bond is a rather weak bond, with the Bi-Si bonds providing
most of the energetic stability.  This can be seen by considering the
electronic structure of a Bi ad-dimer on the clean
surface\footnote{While not a perfect mimic of the Bi dimer in the
  Haiku structure, the smaller cell allows more detailed calculations
  to be made.}.  The electron localisation function
(ELF)\cite{Becke1990} for this ad-dimer is shown in
Figure~\ref{fig:BiDimerELF}(a)\cite{Bowler2005}.  The ELF gives a
quantitative analysis of the bonding properties of a system, with a
value of 0.5 equivalent to the localisation seen in the homogeneous
electron gas and a value of 1.0 equivalent to complete localisation.
The figure shows an isosurface with ELF=0.8 (equivalent to reasonably
strong covalent bonds).  It can be seen that the Si-Si bonds are
rather strong, as expected, and that the Bi-Si bonds are still visible
at this level, though they are weaker than the Si-Si bonds.  The Bi-Bi
bond is not visible (in fact it has a maximum value around 0.7) but
the Bi lone pairs are visible as rather diffuse, spherical objects (by
comparison to the up atoms of the substrate dimers).  This implies
that the Bi-Bi bond is rather weak, and the lone pairs would not be
available for bonding\cite{Bowler2005}.  The energy barrier to
breaking the bond in this Bi dimer has been calculated with DFT to be
0.17\,eV going from over the dimer row to over the trench between rows
and 0.15\,eV going in the other direction, and the barrier is shown in
Fig.~\ref{fig:BiDimerELF}(b).  Having the Bi dimer over the dimer row
is more stable than over the trench between rows by about 0.02\,eV.
This implies that there should be a finite chance of imaging the Haiku
structure with the Bi dimers out-of-phase with the underlying Si dimer
rows.  However, we note that with the energy difference noted above,
the dimer would be over the row 98\% of the time at room temperature, and the
likelihood that a whole line of dimers would flip, will be even smaller.
There are occasional images with the dimers out-of-phase, but these
are rare.

In the next section, details of some elements which will attack the
Bi-Bi bond will be presented, along with others that will not.  Here we
consider the local structure of the bond.  When the Bi-Bi bond is over
the Haiku core, it is almost perfectly relaxed, with bond angles
extremely close to 90$^\circ$.  Insertion of any species into the
Bi-Bi bond will distort these angles (and the angles formed with the
underlying Si atoms), with a potentially large energy penalty.  Thus
the Bi-X-Bi angle should be relatively close to 180$^\circ$ with short
Bi-X bond lengths if an element X is to insert successfully into the
bond. Furthermore, the lone pair on the Bi dimer, which might be
expected to be a strong electron donor, is in fact rather passive, as
it lies some distance below the Fermi level\cite{Bowler2005}. The
implications of this for the chemical reactivity of the Bi nanoline
are discussed below in Section~\ref{sec:reactivity}.

\subsubsection{Strain-induced Enhancement}
\label{sec:el-strain}

The phenomenon of strain-enduced enhancement at low STM bias voltages
has been explored before\cite{Owen1995,Schofield2004}; we reproduce
the arguments here for convenience\cite{Owen2003}.  The Si(001) surface shows a dimer
reconstruction which provides three bonds per surface Si atom
(compared to two bonds per atom for the bulk terminated surface), but
which pulls the bond angles and distances away from the ideal
tetrahedral values.  The net result is that the local band gap is
reduced (in simple chemical terms, the hybridisation towards sp$^3$ is
decreased, leading to a decreased splitting between the resultant
hybrid bonds).  If the surface dimers are then further distorted,
their local states will move closer to the Fermi level.  The
single missing dimer defect (1DV) has second layer atoms bonding
across the defect which distorts the dimers either side of the defect
(though lowering the energy because the second layer atoms have no
dangling bonds).  This distortion in turn causes the states near the
Fermi level on the neighbouring dimers to move towards the Fermi
level, leading to a brightening or enhancement at low biases.

\begin{figure}
  \centering
  \includegraphics[width=\columnwidth]{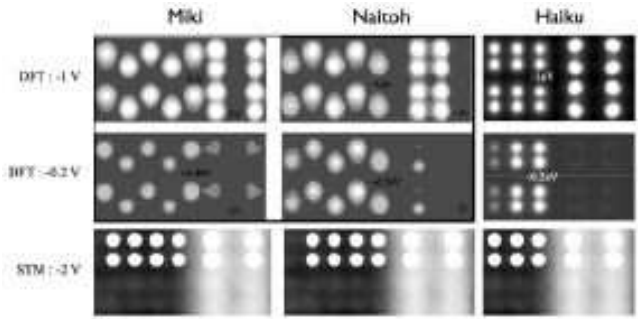}
  \caption{A comparison of simulated STM for the 3 proposed models.
   An STM image of the Bi nanoline has been aligned with
    the centre of the calculated nanoline for each structure. For the
    Miki model, the background Si dimers are offset relative to the
    experimental image, while for the Naitoh model, the Bi dimer
    spacing is much too small. For the Haiku model, both of these
    aspects fit very well. The change in contrast of the Bi nanoline,
    and the dimers adjacent to it, at low biases is discussed in the
    text.  The data for the Miki and Naitoh models are reprinted with
    permission from Ref~\protect\cite{Srivastava2004}.  Copyright
    (2004) by the American Physical Society.}
  \label{fig:simSTM}
\end{figure}

Low-bias simulated images of the Naitoh and Miki
structures\cite{Miwa2002b,Miwa2002a,Srivastava2004} show that the Si
dimers adjacent to the Naitoh become \emph{darker} than the
surrounding Si dimers as the Fermi level is approached, while those on
the Miki model are unchanged.  By contrast, with the Haiku structure,
the Si dimers immediately adjacent to the nanoline become brighter
closer to the Fermi level\cite{Owen2003,Owen2004}, in good agreement
with the STM images.  A comparison of high and low-bias simulated
images of the 3 structures are shown in Fig.~\ref{fig:simSTM}.
However, the prediction from the Miki model does agree very well with
the appearance suggested from STM images\cite{Miki1999a,Naitoh2001},
in which the shoulder dimers around a 1DV on a Bi-rich surface become
dark at low bias voltages.  On the basis of these calculations, we
suggest that the lack of enhancement at low bias voltages indicates
almost complete relaxation of the strain of the 1DV by the presence of
the Bi dimers in the Miki model.

\begin{figure*} 
\centering
\includegraphics[width=0.9\textwidth]{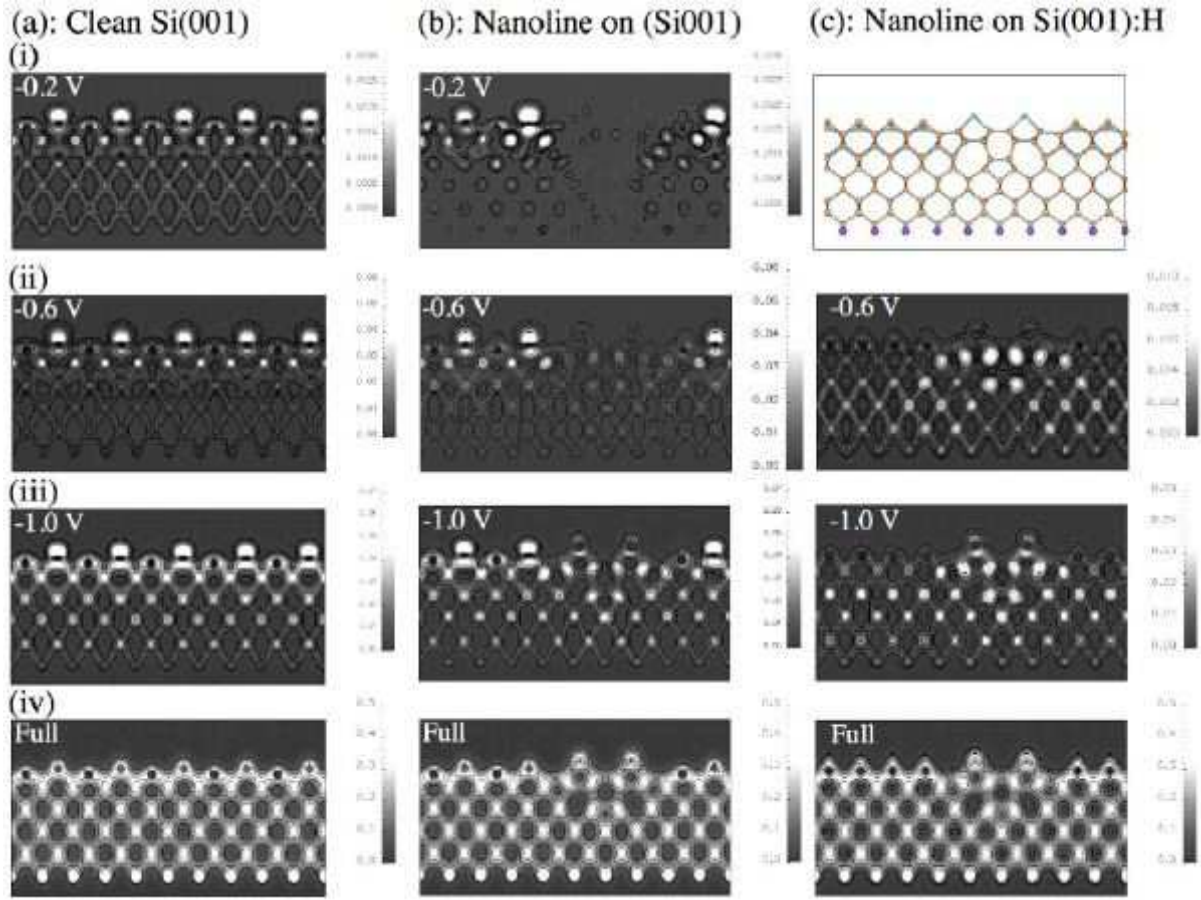} 
\caption{Contour plots of charge density for the states (i) 0.2\,eV,
  (ii) 0.6\,eV and (iii) 1.0\,eV below E$_F$, and (iv) the total
  charge density for: (a) the clean Si(001) surface (first column);
  (b) the Haiku structure on the clean surface (second column), and
  (c) the hydrogenated surface (third column).  The 0.2\,eV image for
  the hydrogenated surface has been replaced with a ball-and-stick
  model of the Haiku structure, as there are no states within 0.2\,eV
  of E$_F$.  (The highest electron density contours are 0.003, 0.060,
  0.070 and 0.500 electrons/cubic \AA in the 1st-4th row respectively
  of columns a,b. The maximum contours in column c are 0.01,0.05,0.50
  electrons/cubic \AA in the 2nd-4th row.)  See text for a full
  discussion.  Reprinted from Surf. Sci. Lett. 527, Owen et al.
  Interaction between strain and electronic structure in the Bi
  nanoline, p.177, (2003), with permission from
  Elsevier.\label{fig:LDOS}}
\end{figure*} 
 
Taking the principle of strain-induced enhancement further, the local
subsurface strain could be determined by cross-sectional STM of the Bi
nanoline, of which there is none, or from LDOS calculations, by
projection of the charge density on a plane perpendicular to
the (001) surface.  We show the projected charge density from GGA
calculations for the clean Si(001) surface, the Haiku structure and
the Haiku structure with hydrogen on the Si(001) surface in
Fig.~\ref{fig:LDOS}\cite{Owen2003}, on the (1$\bar{1}$0) plane, parallel to a Si dimer row.
By comparing in particular the clean Si(001) surface and the clean Haiku structure, the change in
different states induced by the Haiku can be seen.  Beginning with the
clean surface, the localisation of the states close to the Fermi level
in the top few surface layers shows the strain induced by
dimerisation.  In the Haiku structure, these states are even more
strongly localised on the dimers immediately adjacent to the Haiku
structure, which causes their relative enhancement in STM. Meanwhile
the triangular Si substructure of the Haiku, c.f
Fig.~\ref{fig:haikucore}, is somewhat relaxed compared to the clean Si
dimers, although it is strained relative to the hydrogenated dimers
and perfect bulk Si.  This goes some way to explaining the stability
of the subsurface 5-7-5 ring structures, which serve as a highly
effective relief mechanism for the epitaxial stress exerted by
adsorbed Bi.  The absence of states close to the Fermi level suggests
that the nanolines are likely to block surface conduction
perpendicular to the nanoline, and are not likely to act as a
nanowire.

The lack of any states on the Bi near the Fermi level is clearly the
cause of the darkening of the nanoline relative to the Si(001) surface
seen in STM at low voltages.  In the
full charge densities(Fig.~\ref{fig:LDOS}(iv)), the density of states
associated with the Bi dimers, combined with their higher physical
height, explains their relative brightness in STM at higher bias
voltages. In the case of the hydrogenated surface, the Bi dimers have
a similar charge density as in the clean surface, but in this case,
the Si $\pi$-bonds have been eliminated, so the Bi dimer remains
bright in STM at all biases\cite{Owen2002a}.

\subsection{Chemical Properties}
\label{sec:reactivity}
As described above in more detail, the bonding orbitals in the Bi
dimers have essentially a p$^{3}$ bonding character, while the lone
pair is a diffuse s-type orbital, some 10 eV below the Fermi
level\cite{Bowler2000}.  Thus the Bi dimer is in a very stable
electronic state, and is not expected to be very reactive.  Moreover,
this low-energy configuration is dependent upon the maintenance of
bond angles close to 90$^{\circ}$, which will have an effect on the
energetics of any insertion into the Bi-Bi dimer bond. However, the
low-bias images shown above reveal that although the nanoline appears
clean at high biases, some contrast between different dimers is
visible at lower biases, suggesting that some chemical attack has
occurred.
 
\subsubsection{Reactivity of the Bi dimer}
\label{sec:reactivity-bi-dimer}

We have proposed that the Bi nanolines, although semiconducting
themselves, might be used as templates for deposition of other
materials, such as metals, active molecules, nanoparticles
etc.\cite{Owen2002a,Owen2005c}. In order for the nanoline to have
utility in this role, it must be possible to mask the substrate around
the nanoline, so that adsorbants stick preferentially to the nanoline.
It is also important, for any nano-electronics applications, to have
some method for rendering the substrate insulating after growth of the
nanoline. The chemical reactivity of the Bi dimers, and the nanoline
as a whole, is therefore crucial to any implementation of this
concept.  Atomic hydrogen has been used as a mask on the bare Si(001)
surface for STM lithography
experiments\cite{Hashizume1996,Shen1997,Sakurai2000}, in which an STM
tip is used to remove lines or areas of hydrogen from the 
surface, and thus provide areas onto which metal will adsorb, as in
Fig.~\ref{fig:Hashizume}.  However, the writing process is slow, and
lacks scalability. Use of the Bi nanolines as templates would remove
the need for this writing, while providing long, and more uniform
patterns. Hydrogen is therefore a natural species to use as a mask
around a Bi nanoline.
 
One of the most favourable properties of silicon as an electronics
material is the high quality of its oxide film. Oxidation of the
silicon around the Bi nanoline is therefore a natural way in which to
isolate the nanoline electrically from the substrate.  The usual
method of oxidation of silicon is by exposure to molecular oxygen or
to water at elevated temperatures, but this only gives
microelectronics-quality oxide when grown at ca.800$^{\circ}$C. Since the
Bi nanoline is not stable above 600$^{\circ}$C, ozone was considered
as an alternative for low-temperature oxidation. Ozone preferentially
attacks the backbonds of the surface Si dimers, forming a stable
oxide, even at room temperature\cite{Nakamura1999}.  More recently,
the need for high-k dielectrics has driven the development of silicon
nitride and oxynitride as alternatives to silicon dioxide.
Nitridation is usually done using ammonia, which breaks up on contact
with the clean silicon surface above 200 K\cite{Rignanese2000}.  As an
added bonus, when ammonia reacts with silicon, the nitrogen moves
below the surface, but the hydrogen remains on the surface,
passivating it. It can therefore play a dual role, as insulating
reagent, and as masking material. The reaction of these species with
the Bi nanoline are discussed in the sections below.

\subsubsection{Hydrogen}
\label{sec:hydrogen}

\begin{figure}
\includegraphics[width=\columnwidth]{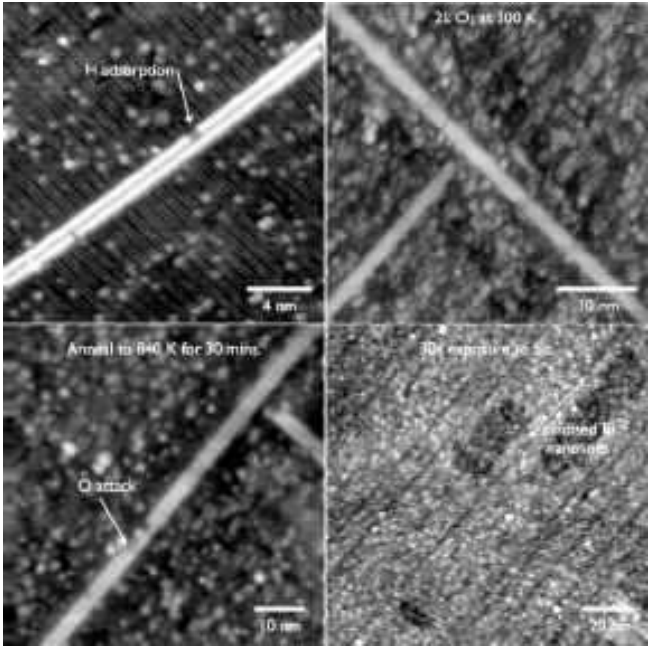}
\caption{a: A filled-states atomic-resolution image of a H-terminated
  surface, showing a section of a nanoline. The imaging conditions
  here are -2.5~V, 0.1~nA. (b): Bi nanolines after exposure to 2L
  O$_{3}$ at 300~K. There is virtually no attack on the nanolines,
  while the substrate has been oxidised. (c): After exposure to 150L
  O$_{2}$ at 700~K and an anneal to 840 K, some gaps appear in the
  nanoline, but most of it is unaffected. (d): Exposure of a
  H-terminated surface with Bi nanolines to air results in complete
  oxidation of the nanolines, which now show up dark relative to the
  H-terminated Si.\label{fig:Bihydrogenoxygen}}
\end{figure}

The Bi nanoline has been shown to be almost completely inert to attack
by atomic hydrogen\cite{Naitoh2000,Owen2002a,Owen2005b}.
Fig.\ref{fig:Bihydrogenoxygen}(a) shows a close-up image of a nanoline
on a Si(001):H monohydride surface. In this image, nanolines were
formed at 590$^{\circ}$C, and the surface was cooled to 330$^{\circ}$C,
the surface was saturated with atomic H. The surface has the
monohydride phase, with the Si-Si $\sigma$-bond intact, and the H
atoms saturating the dangling bonds.  Thus each Si-H dimer is resolved
as a pair of dots. Some white dots on the Si surface may be holes in
the monohydride layer, stray bismuth atoms, or adsorbed -OH resulting
from water impurity in the hydrogen. While the nanoline is almost
completely unreacted, one dimer has become dark. This sort of defect
is extremely rare. The preferential adsorption of the H onto the Si is
consistent with DFT-LDA modelling, which found that H on a Bi dimer is
3~eV worse than H on a Si dimer. There is therefore a strong driving
force for any H which does adsorb onto the Bi nanoline to diffuse off
it, and onto the Si.

\subsubsection{Oxygen} \label{sec:oxygen}

The oxidation properties of a monolayer of Bi on Si(001)
shows an unusual changeover in behaviour with oxygen exposure. An
EELS/Auger study\cite{Koval1997} found that for small exposures, the
reactivity of a Bi overlayer on Si(001) is 100 times less than the
reactivity of oxygen with the bare silicon surface. However, for large
exposures, above ca. ${3\times10^5}$ L O$_{2}$, the oxygen Auger
intensity continues to increase for a Bi/Si(001) surface, at a point
where the signal for the
clean Si(001) surface has saturated. Thus for very large
exposures, the Bi promotes oxidation of the silicon. Moreover, the
stoichiometry of the oxide formed on the Bi/Si(001) surface is closer
to SiO$_{2}$, which would imply better insulating properties.
Semiempirical calculations found that insertion of oxygen into the
Bi-Bi bond is energetically much less favourable than for Sb or As
(0.32 eV vs ca. 4.7 eV). Moreover, insertion into a Bi-Si bond or a
Si-Si bond is much more favourable (4.5 eV and 5.5 eV respectively), so
that preferential attack of the silicon is expected.

The reactions of molecular oxygen and ozone with the Bi nanoline show
a similar changeover in behaviour\cite{Owen2002a}. The surface was
exposed to ozone at room temperature and, for comparison, to molecular
oxygen at 400$^{\circ}$C. Both surfaces were annealed up to
570$^{\circ}$C, near the stability limit of the Bi nanoline.  The
room-temperature ozone experiments, using exposures up to 20 L of
O$_3$, revealed that the silicon background could be saturated without
visible damage to the nanoline. In large-scale images, the Bi nanoline
appears to be completely unaffected, as shown in Fig.
\ref{fig:Bihydrogenoxygen}(b). However, in small-scale images, and by
varying the imaging bias, some subtle differences between the
appearance of various dimers in the nanoline could be
seen\cite{Owen2002a}.  This variation may be due to attack of Bi-Si
backbonds, as found to be favourable in MNDO
simulations\cite{Afanasieva2002}, which would leave the Bi-Bi dimer
intact, but perturb its electronic structure somewhat. Annealing this
surface introduced some defects into the nanoline. After annealing at
450$^{\circ}$C, gaps appear in the nanoline, which comprise 1-2 unit
cells. This trend continued with a further anneal to 570$^{\circ}$C,
as shown in Fig. \ref{fig:Bihydrogenoxygen}(c).  Such piecemeal damage
may be explained by the observation that the activation barrier for Bi
to desorb from oxidised Si has been found to be 0.7 eV lower than for
unoxidised Si\cite{Koval1997}. The small contrast changes, which were
seen after ozone exposure at room temperature may therefore indicate
reaction with Bi-Si, or Si-Si backbonds around the nanoline, so making
that unit cell more susceptible to evaporation during high-temperature
annealing.  Molecular oxygen, when adsorbed at 400$^{\circ}$C, behaved
in a very similar way to ozone. Again there was preferential attack of
the silicon, leaving the nanoline mostly intact, but also gaps
appeared in the Bi nanoline after annealing for about one
hour\cite{Owen2002a}.

While the Bi nanoline is remarkably resistant to oxidation in UHV, it
quickly oxidises when exposed to air, as shown by the following simple
experiment. A H-terminated Si(001) surface with Bi nanolines was taken
out of the imaging chamber into the UHV system loadlock, which was
then vented to atmosphere for 30s, and then immediately pumped back
down to UHV. The resulting surface is shown in
Fig.~\ref{fig:Bihydrogenoxygen}(d).  The nanolines, which previously
were bright at all biases relative to the H-terminated background, are
now black lines, indicating that they have suffered significant
attack. This agrees well with the previous study, which found that for
very large exposures to oxygen, the Bi dimers promoted oxidation of
the underlying silicon.

\subsubsection{Ammonia}
\label{sec:ammonia}

\begin{figure}
  \includegraphics[width=\columnwidth]{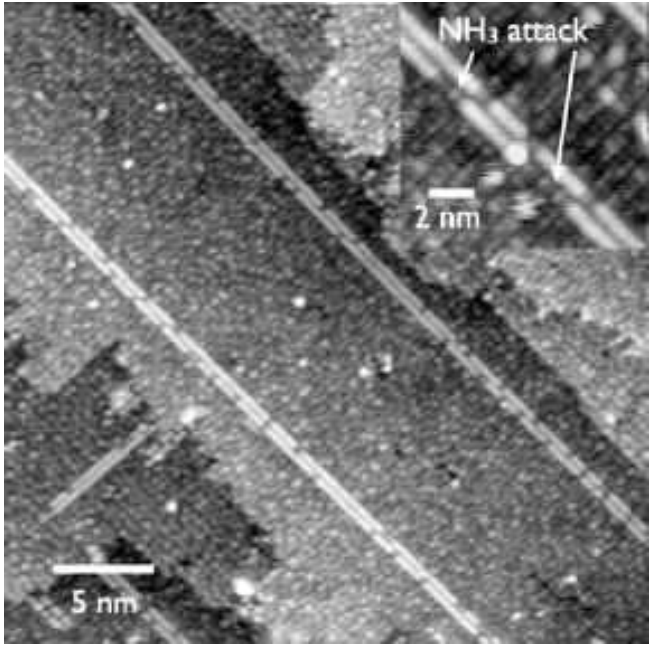}
  \caption{A 64 nm$\times$64 nm STM image of the ammonia-terminated
    Bi:Si(001) surface. The background Si is completely terminated by
    a mixture of H and NH$_{2}$ groups. Dark patches on the Bi
    nanoline reveal some attack by ammonia. The inset shows that
    reaction occures within the space of a single dimer, and the
    resulting feature is asymmetric, much like the termination of the
    background Si. }
  \label{fig:Biammonia}
\end{figure}

The adsorption of ammonia onto Si(001) as a precursor to silicon
nitride or oxynitride growth has been studied by a variety of
experimental and theoretical techniques, which concluded that at room
temperature, the ammonia dissociated on the surface, forming NH$_{2}$
and H groups\cite{Rignanese2000}. We have found that ammonia attacks
the silicon preferentially at room temperature. An image of the
Bi:Si(001) surface with ammonia termination is shown in
Fig.\ref{fig:Biammonia}. Within each dimer, the light grey end has
been identifed with the NH$_{2}$ group, while the dark grey end is the
H atom\cite{Owen2005a}. The apparent height difference in STM matches
the physical height difference between these two species.  There is
some ordering of the NH$_{2}$ groups, such that zigzag patterns and
straight lines are seen running along the dimer rows.

Reaction of ammonia with the Bi nanoline might be expected to proceed
by the initial formation of a hydrogen bond between an ammonia H atom
and the Bi lone pair, followed by insertion into a Bi-Bi or Bi-Si
bond. However, as discussed above, the lone pair on the Bi dimer is
essentially inert\cite{Bowler2005}.  During experiments on a surface
which was pre-exposed to atomic H to saturate the Si dangling bonds,
for exposures of up to 70L of ammonia at substrate temperatures of up
to 500K, the Bi nanoline was largely unaffected.  However, adsorption
of ammonia onto a Bi nanoline surface where the Si dimers are not
passivated does result in some reaction with the nanoline, even at
room temperature. This difference may indicate that ammonia reacts
first with the silicon, forming active NH$_{2}$ species, which are
mobile and can attack the Bi dimers.  Calculations suggest that it is
energetically favourable for an NH$_{x}$ group to insert into the
Bi-Bi dimer bond, and form a symmetrical Bi-NH-Bi feature. However,
STM images of reacted Bi dimers shows an asymmetric feature, not
unlike the feature seen on background Si dimers. Examples may be seen
in the inset to Fig.~\ref{fig:Biammonia}. Nevertheless, the
probability of attack by ammonia is low, providing the promise that
ammonia can be used to grow silicon nitride without damage to the
nanoline.

\subsubsection{Metals}
\label{sec:metals}

\begin{figure}
   \includegraphics[width=\columnwidth]{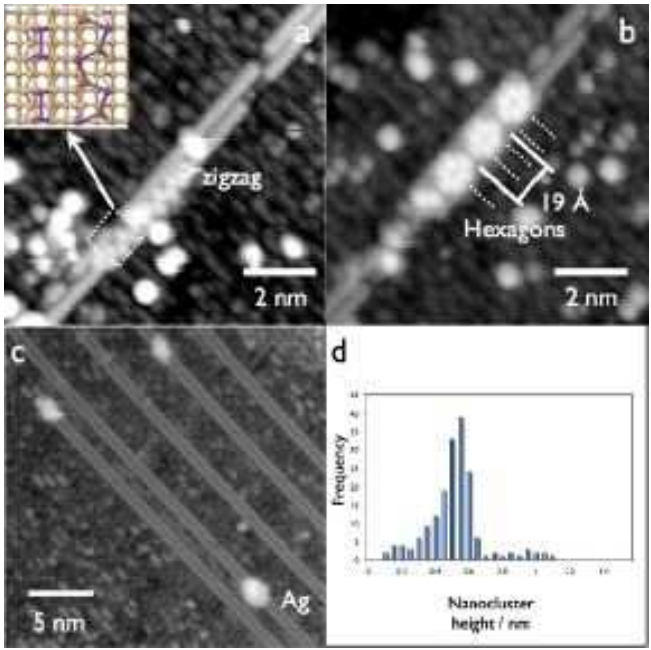}
   \caption{(a): Indium reacts with the Bi nanoline to form a zigzag
     feature. (b): A two-layer In island shows hexagonal features,
     which are 19\AA\ or 2.5 dimer rows wide, revealing an epitaxial
     relationship with the Bi nanoline. (c): Isolated 0.5~nm Ag
     nanoclusters form after deposition of Ag atoms at room
     temperature onto the Bi nanoline surface. (d): Height Analysis
     reveals a strongly-peaked size distribution, with the peak around
     0.5~nm, which may correspond to a magic cluster of 13 atoms. }
  \label{fig:metal}
\end{figure}

Metal deposition onto the Bi nanolines is currently under
investigation.  Deposition of Ag onto Bi nanolines on the clean
Si(001) surface\cite{Itoh2005} reveals a strong preference for
adsorption onto the Si, leaving the Bi nanoline clean (in complete
contrast to the behaviour of ErSi$_2$ NWs on Si(001) discussed above
in Sec.~\ref{sec:reactivity-1}).  However, the passivity of the Bi
nanolines to atomic hydrogen and ammonia allows these species to be
used as nanoscale masks. Ag is known not to adsorb onto H-terminated
Si(001)\cite{Sakurai2000} as shown in Fig.~\ref{fig:Hashizume}, while In 
completely dewets from the Si(001)
surface, forming large droplets when H is adsorbed onto an In-covered
surface\cite{Zotov1997}.  In all cases, following metal deposition
onto an ammonia-terminated Si(001) with Bi nanolines, preferential
adsorption onto the Bi nanoline (or onto background Bi) was seen,
indicating that the ammonia-terminated surface is very effective as a
mask against metal deposition.  The reaction of the metal with the
nanoline template is more complex, however.  Two types of behaviour
have been seen with different metals, which may be described as
wetting and non-wetting behaviour. Wetting occurs when the interaction
of the metal with the nanoline is stronger than the interaction of the
metal with itself.  Nonwetting occurs when the interaction of the
metal with itself is stronger. Examples of each type of behaviour for
different metals, In and Ag, are shown
in Fig.~\ref{fig:metal}. Adsorption of In results in significant
intermixing of In and Bi, forming a zigzag island structure, which is
thought to result from a chain of In and Bi atoms, as shown in the
inset to Fig.~\ref{fig:metal}(a). Second-layer islands have a distinctive
bright hexagonal feature, which is 19~\AA\ or 2.5 dimer rows wide.
This rational relationship between the In island dimension and the
underlying Si unit cell distance demonstrates that these In islands
grow \emph{epitaxially} on the nanoline template, raising the
possibility of the growth of long single-crystal nanowires on the
Si(001) surface. By contrast, adsorption of Ag results in the
formation of small clusters of Ag, around 0.5~nm in height, at a very
early stage of deposition, as shown in Fig.~\ref{fig:metal}(c).
Further deposition at room temperature results in an increase of the
number density of these nanoclusters, but the modal size increases
only slightly from 0.5~nm to 0.6~nm, with a strongly peaked height
distribution.  This behaviour may be thought of as non-wetting
behaviour.

\subsection{Burial}
\label{sec:burial}

While the Bi nanoline has a number of interesting properties, for any
nano-electronics device application it must be possible to passivate
them against atmospheric attack. One way in which to do this is to
bury them in a layer of epitaxial silicon. Moreover, the burial of the
Bi atoms may lead to their acting as a 2D delta-doping layer, with
close to atomic-layer precision in the depth placement.  However, as
with other nanostructures, for example burial of InAs/GaAs quantum
dots in GaAs, the process of burial is likely to change the shape and
detailed structure

\begin{figure}
\centering
  \includegraphics[width=0.9\columnwidth]{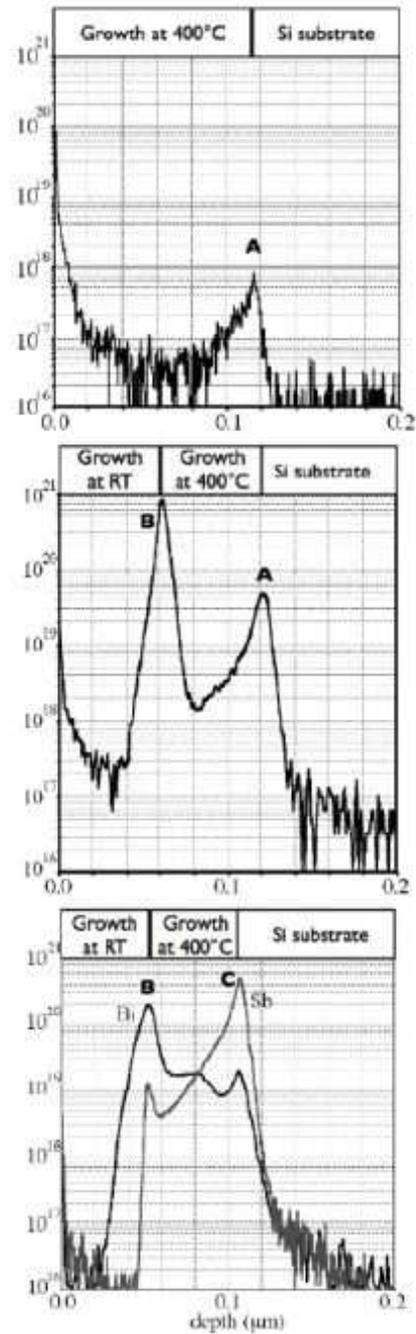}
  \caption{(a): A SIMS profile of a Bi nanoline surface, formed at the position marked `A' 
    and then overgrown with Si at a substrate temperature of 400$^{\circ}$C. Most of the Bi
    has segregated to the surface, and the nanolines have been
    destroyed. (b): The Bi nanoline surface (A) has
    been capped with a monolayer of Bi at 400$^{\circ}$C, before
    overgrowth. After overgrowth at 400$^{\circ}$C up to position `B', the 
    growth layer was buried in amorphous Si by further deposition of Si at room
    temperature. In this case, nearly all the Bi from the Bi nanolines
    remains at the original surface position, `A'. (c): As (b), except that
    the capping layer material was Sb, in place of Bi. Here most of
    the Bi has segregated to the growth surface (B), leaving Sb at the
    original surface (C), indicating that the Bi has exchanged with Sb
    from the capping layer.}
  \label{fig:SIMS}
\end{figure}

Simple overgrowth of a surface containing Bi nanolines results in the
destruction of the nanolines\cite{Miki1999c,Yagi2005}. As shown in the
Secondary Ion Mass Spectrometry(SIMS) profile in
Fig.~\ref{fig:SIMS}(a), after overgrowth of Si the vast majority of
the Bi has segregated to the surface of the grown film, although there
is a small peak in the Bi concentration at the original surface
position. Furthermore, X-ray analysis\cite{Yagi2005} shows that the
1-D character of the nanoline is lost after capping, even by an
amorphous overlayer deposited at room temperature.  The mechanism by
which the Bi floats to the surface during growth is thought to involve
exchange of subsurface Bi atoms with surface Si atoms, as Bi is an
effective surfactant during growth. In order to prevent this, a
further surfactant layer of Bi can be deposited around and over the Bi
nanolines, before Si overgrowth\cite{Miki1999c}. In this case, the
exchange mechanism is inoperative, as the second-layer Bi in the
nanolines will not exchange with first-layer Bi, and so it is possible
to block the segregation of the Bi nanolines to the growth surface. In
the SIMS profile in Fig.~\ref{fig:SIMS}(b), a thin crystalline Si
overlayer has been grown at 400$^{\circ}$C after deposition of a Bi
layer, and this has been capped by depositing further Si at room
temperature, so as to bury the growth surface in amorphous Si. In this
case there, are two peaks in the SIMS profile. `A' refers to the
Bi density at the original Bi nanoline surface, while `B' refers
to the surface of the crystalline overgrowth. The surface of the
overlayer has approximately 1 ML of Bi on it, from which the quantity
of buried Bi can be determined. This is approximately 5\% ML, which is
typical for a Bi nanoline surface.  Thus, using the Bi surfactant
layer, most of the Bi nanolines have been preserved at the original
surface.  In order to gain more information about the exchange
process, a surfactant layer of Sb was also used, instead of the
surfactant layer of Bi. The resulting SIMS profile is shown in
Fig.~\ref{fig:SIMS}(c).  In this case, there is a large peak of Sb at
the original nanoline surface position `B', while a considerable fraction
of the Bi has segregated to the surface of the overlayer `C'. This result
indicates that there is considerable exchange between the Sb
surfactant layer and the Bi nanolines, and by this method, a layer of
Sb has been successfully buried, with a sharp peak at the original
nanoline surface.

\begin{figure}
\begin{center}
  \includegraphics[width=0.9\columnwidth]{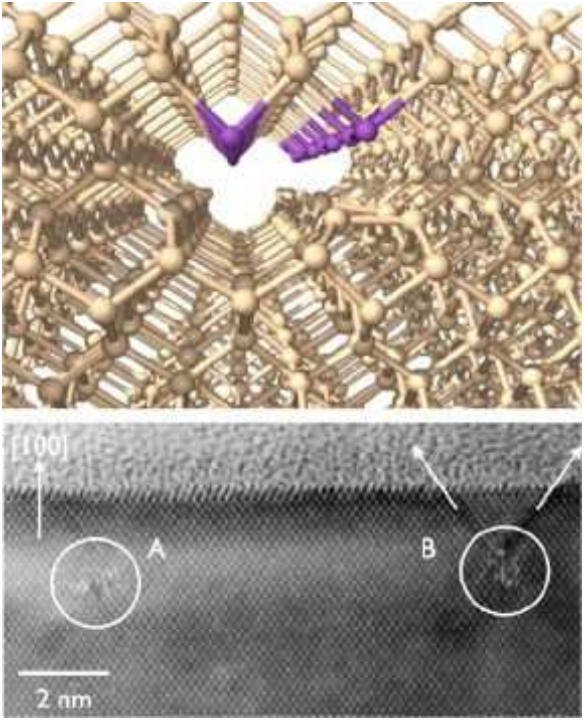}
  \caption{(a): A ball-and-stick model of a proposed structure for a
    buried Bi nanoline. The haiku structure has reconstructed into a
    2DV-like trench, with the Bi dimers attached to Si overhead. (b):
    A cross-sectional TEM image of Bi nanolines buried in a Si
    epitaxial layer. The one on the left is defect free, while the one
    on the right has triggered dislocations.}
  \label{fig:buriedhaiku}
  \end{center}
\end{figure}

While this exchange process is successful at burying the Bi atoms, it
does not provide any information about the structure of the buried Bi.
For device applications, it is important to retain the 1-D character
of the original nanoline. It is also of interest whether the Haiku
structure will be preserved.  STM cannot observe subsurface features,
and instead the structure of the buried Bi nanolines was studied using
an X-ray method\cite{Sakata2004}. The results of this
analysis\cite{Sakata2005} indicate that two important elements of the
Bi nanoline structure have been preserved: the 1-D character, and the
dimerisation of the Bi atoms. Tightbinding and DFT calculations of the
buried Bi nanoline structure found that the Haiku structure was not
stable, and broke down into a hollow 1-D structure with Bi dimers
bonding to the silicon grown \emph{over} the nanoline, and the Si
substructure reconstructing into something similar to a 2DV surface
defect.  A ball-and-stick model of this proposed structure is shown in
Fig.~\ref{fig:buriedhaiku}(a).  It has significantly lower energy per
Bi atom than other proposed structures, including substitutional and
interstitial Bi defects.  This structure may be compared to the
cross-sectional TEM image of buried Bi nanolines in
Fig.~\ref{fig:buriedhaiku}(b). Two nanolines are present within the
TEM image. The one on the left has strain associated with it, and a
dark core, which may match with the proposed structure. The one on the
right has triggered misfit dislocations along \{111\} planes, c.f.
Fig.~\ref{fig:haikucore}. These dislocations, which indicate the
presence of a large local stress, may suggest that the Haiku
substructure of the nanoline has survived overgrowth.  A recent X-ray
Standing Wave (XSW) study of a Bi nanoline sample which had been
capped with amorphous Si at room temperature\cite{Saito2003} has cast
doubt on the validity of the Haiku model.  However, more detailed
analysis of lines buried with amorphous silicon\cite{Sakata2005}
showed that the dimerisation and one dimensional character of the
nanolines is destroyed by burial in amorphous silicon (and, as
discussed above, burial in crystalline silicon changes the Haiku
reconstruction while leaving the one dimensional, dimerised character
of the nanolines unaffected).  Furthermore, the presence of a
significant quantity of Bi at the growth surface may have an effect on
the results obtained, unless it is removed\cite{Yagi2005}.

\section{Conclusions}
\label{sec:conclusions}

We have given a broad overview of the different material systems and
experimental methods which result in self-assembled 1D nanostructures on
semiconductor surfaces, particularly the technologically important
Si(001) surface. There are many different self-organising systems which 
result in 1D nanoscale features on the surface of Si(001) and Si(111).
Aside from the Pt/Ge(001) system, none of these approach the length, and
degree of perfection, of the rare-earth silicide wires and the Bi nanolines.
The metal/Si(111) nanowire systems will tend to form 3 equivalent domains,
limiting the long-range order of the nanowires, while the methods which rely on
step-edge adsorption are limited by the spacing of step kinks (though
this can be controlled with careful sample preparation).
The rare-earth silicide nanowire family are not nanowires \textit{per
  se} (in that they are a metastable state, not seen in a bulk structure), but are
better regarded as a conventional heteroepitaxial system in which the lattice mismatch
is anisotropic. Even so, while many different rare earths superficially appear to behave in
a similar fashion, review of the literature indicates that in fact several different nucleation
and growth mechanisms are responsible for the formation of nanowires within this family,
both kinetic and thermodynamic.  The recent observation of nanowires with an orthorhombic
crystal structure\cite{Harrison2005b}, rather than the hexagonal structure previously reported, 
suggests that there is much more to be learnt about this family of materials.
Experimental studies of all these systems has so far been largely 
confined to STM, apart from X-ray and TEM studies of buried Bi nanolines, and TEM
of the endotaxial RE nanowires. A much greater understanding of the 
physical, chemical and electronic properties of these systems, would come from
other experimental techniques, such as optical and vibrational spectroscopic techniques.
(For example, the strained Haiku core of the Bi nanoline may have a signature bond
vibration frequency.)

The main focus of this review has been the Bi nanoline system. We have described the 
physical and electronic structure of the Bi nanoline, its reactivity 
with different materials, and the surface phenomena associated with it.  
Although there has been no direct observation of the core structure of the Haiku model, 
it provides a natural explanation for all the observed experimental phenomena, and
is well-supported by experimental and theoretical data. The 
fundamental source of many of the properties of the Bi nanoline is the
subsurface core of Si which has recrystallized into a hexagonal phase,
Lonsdaleite (known from meteorites). The formation mechanism of this core structure 
remains unknown. A significant activation barrier is expected, given the high temperature and 
long incubation time required to form the nanolines, and the co-existence 
of the nanolines with the Miki structure, which is less stable, but  
kinetically easy to form. While mechanisms proposed thus far have 
concentrated on the twisting of pairs of atoms, an alternative may well 
be a stacking-fault-like shift of the entire core from the cubic phase to the 
hexagonal phase.  
Given their structure, it is clear that the Bi nanolines are not a conventional 
heteroepitaxial system like the silicide family.  There is no variation in width, 
either with annealing temperature or time -- only the number density of nanolines
and the length changes with further deposition and annealing. 
The nanolines are only ever one unit cell wide, although clusters of nanolines 
are sometimes seen. The nanolines are not the result of simple adsorption onto
symmetry-breaking features of the surface, such as step edges, or
linear defects.  Nor are they a periodic reconstruction of the
surface, as in the Pt/Ge wires\cite{Gurlu2003}. They might best be described 
as an adsorbate-stabilized surface recrystallization.

The chemical properties of the nanoline stem from the behaviour of the
Bi dimer with its preference for p$^{3}$ bonding and an inert s-type
lone pair, rather than the sp$^{3}$ hybridization of Si dimers. 
Although the Bi-Bi bond is weak, it is also passive, and
only In shows any significant reaction with it, probably because the
zigzag structure produced allows the Bi atoms to retain their
preferred 90$^{\circ}$ bond angles. Hydrogen does not attack the Bi
nanoline, while ozone will oxidise around the nanoline at room
temperature, leaving it intact. Likewise ammonia preferentially attacks the
silicon, although some damage to the nanoline occurs at small fluxes.
 
The determination of both the physical and electronic structure of the 
nanoline has demonstrated the importance of a close interaction 
between experiment and theory. We note that all the recent publications 
on the Bi nanoline have been a collaboration between the two. 
The method of tightbinding has proved to be a valuable tool for quickly 
searching through a large variety of possible structures, as in for instance the Haiku 
structure, or for relaxing a large number of atoms, as in modelling of the 
nanoline kink energies. At the same time, the tightbinding method retains 
a quantum mechanical description of the energetics, so that the calculated 
energies are in most cases in reasonable agreement with those calculated using DFT.
The Bi nanoline has benefitted from an unusual degree of theoretical investigation.
A similar level of theoretical work for the rare-earth nanowires is likely to
deepen greatly the understanding of the different nucleation and growth 
phenomena that have been identified in that family of systems.

With a significant body of work devoted to the fabrication and structure of the
nanoline systems, attention should be paid to their possible applications.
The length, straightness and perfection of self-assembled nanolines 
lend themselves to use as a nanoscale template. The most obvious
candidate systems are the rare-earth nanowires, which are themselves metallic, 
so that an array of parallel nanowires would readily form the contacts for a 
nanoelectronic device, perhaps using the ``crossbar'' architecture. The Er 
silicides appear also to act as preferential adsorption sites for metals such as Pt, 
so that the properties of these contacts is not limited to that of the nanowires 
themselves. However, the growth of these nanowires on a semiconducting 
substrate, with no indication as yet that the substrate could
subsequently be oxidised,  might limit their application in nanoelectronics 
(for which a high-quality insulating substrate is crucial).
For the Bi nanolines, oxidation of the surrounding
silicon with ozone, followed by passivation of the surface with
hydrogen, or nitridation using ammonia, is a plausible route for the fabrication of
a single-nm 1D template on an insulating surface, considerably smaller than any of the
rare-earth nanowires, and well within the size range where quantum scale effects
might be expected to occur. By the deposition of metals, 
active molecules, nanoparticles, or other species onto this template, the 
fabrication of nanowires, or arrays of nanoparticles with interesting optical, 
electronic or magnetic properties could be achieved, with atomic precision.  
This area is being actively pursued by the authors.


\begin{acknowledgement}
  We are happy to acknowledge useful discussions with Chigusa Ohbuchi
  and Wataru Yashiro, as well as permission from various authors to
  reproduce figures as noted in the text.  This study was performed
  through Special Coordination Funds for Promoting Science and
  Technology from the MEXT, Japan (ICYS and Active Atom-Wire
  Interconnects).  DRB is funded by the Royal Society.
\end{acknowledgement}

\bibliographystyle{jmatsci}

\end{document}